\documentstyle[astrobib,psfig]{mn-ab}

\title[The proximity effect in a close group of QSOs]{The proximity effect in a close group of QSOs}

\author[J. Liske et al.]{J.~Liske$^1$\thanks{E-mail: 
	jol@phys.unsw.edu.au} and G.~M.~Williger$^2$\\ 
	$^1$School of Physics, University of New South Wales, Sydney 2052,
	Australia\\ 
	$^2$NASA Goddard Space Flight Center, Greenbelt, Maryland 20771, USA}

\date{Accepted
...... Received .....}

\pagerange{\pageref{firstpage}--\pageref{lastpage}}

\pubyear{2000}

\newcommand{\lya}{\mbox{Ly$\alpha$}}
\newcommand{\be}{\begin{equation}}
\newcommand{\ee}{\end{equation}}
\newcommand{\bea}{\begin{eqnarray}}
\newcommand{\eea}{\end{eqnarray}}
\newcommand{\eref}[1]{(\ref{#1})}
\renewcommand{\d}{{\rm d}}
\newcommand{\ion}[2]{#1$\,${\sc #2}}
\newcommand{\kms}{~km~s$^{-1}$}

\begin{document}

\label{firstpage}
\maketitle

\begin{abstract}
We present an analysis of the proximity effect in a sample of ten
$2\,$\AA\ resolution QSO spectra of the \lya\ forest at $\langle z
\rangle = 2.9$. Rather than investigating variations in the number
density of individual absorption lines we employ a novel technique
that is based on the statistics of the transmitted flux itself. We
confirm the existence of the proximity effect at the $> 99$ per cent
confidence level. We derive a value for the mean intensity of the
extragalactic background radiation at the Lyman limit of $J =
3.5^{+3.5}_{-1.3} \times 10^{-22}$ ergs s$^{-1}$ cm$^{-2}$ Hz$^{-1}$
sr$^{-1}$. This value assumes that QSO redshifts measured from high
ionization lines differ from the true systemic redshifts by $\Delta v
\approx 800$\kms. We find evidence at a level of $2.6\sigma$ that the
significance of the proximity effect is correlated with QSO Lyman
limit luminosity. Allowing for {\em known} QSO variability the
significance of the correlation reduces to $2.1\sigma$.

The QSOs form a close group on the sky and the sample is thus well
suited for an investigation of the foreground proximity effect, where
the \lya\ forest of a background QSO is influenced by the UV radiation
from a nearby foreground QSO. From the complete sample we find no
evidence for the existence of this effect, implying either that $J >
20 \times 10^{-22}$ ergs s$^{-1}$ cm$^{-2}$ Hz$^{-1}$ sr$^{-1}$ or
that QSOs emit at least a factor of $1.4$ less ionizing radiation in
the plane of the sky than along the line of sight to Earth. We do
however find one counter-example. Our sample includes the fortunate
constellation of a foreground QSO surrounded by four nearby background
QSOs. These four spectra all show underdense absorption within $\pm
3000$\kms\ of the redshift of the foreground QSO.
\end{abstract}


\section{Introduction}
The study of many physical processes at high redshift requires
knowledge of the intensity of the UV background radiation, $J$. For
example, it is thought that the \lya\ forest in QSO absorption spectra
is caused by highly photo-ionized gas and thus an estimate of the
total mass content of the intergalactic medium (IGM) depends on $J$
\cite{Rauch97,Weinberg97}. It is also one of the parameters that
define the environment in which galaxies form (e.g.\ \citeNP{Susa00}),
and its value and evolution provide important constraints on the
objects believed to be the origin of the background (e.g.\
\citeNP{Bechtold87}; \citeNP{Haardt96}; \citeNP{Fardal98}). At high
redshift, the background is often measured from the proximity effect,
i.e.\ the observed underdensity of \lya\ forest absorption lines in
the vicinity of background QSOs.

Nearly $20$ years ago \citeN{Carswell82} first noted that the mean
density of \lya\ absorption lines seemed to increase with redshift
when comparing the spectra of several different QSOs and yet decrease
along individual lines of sight. \citeN{Murdoch86} confirmed this
`inverse effect', established that it was confined to the vicinity of
the QSO (`proximity effect') and offered two possible
explanations: i)~the absorbers near a QSO may be too small to fully
cover the continuum emitting region or ii)~absorbers in the vicinity
of a QSO may be more highly ionized than elsewhere due to the QSO's UV
radiation.

In a seminal paper \citeN{Bajtlik88} (hereafter BDO) developed a
quantitative ionization model and, for the first time, measured $J$
from the observed underdense absorption near $19$ QSOs and their
observed luminosities. In the most comprehensive intermediate
resolution study to date \citeN{Scott00} analysed a sample of $74$
spectra. Like \citeN{Bechtold94}, they divided their sample into low
and high-luminosity subsamples and found that the relative deficit of
absorption lines within $1.5~h^{-1}_{75}$~Mpc of the background QSOs
was more significant in the latter.

\citeN{Cooke97} studied the proximity effect at high spectral
resolution. In their thorough analysis of $11$ spectra they gave a
detailed account of the various statistical and systematic errors and
biases involved in the measurement of $J$ from the proximity
effect. Like previous authors they found no evidence to suggest that
the background intensity evolved with redshift.

\begin{figure}
\psfig{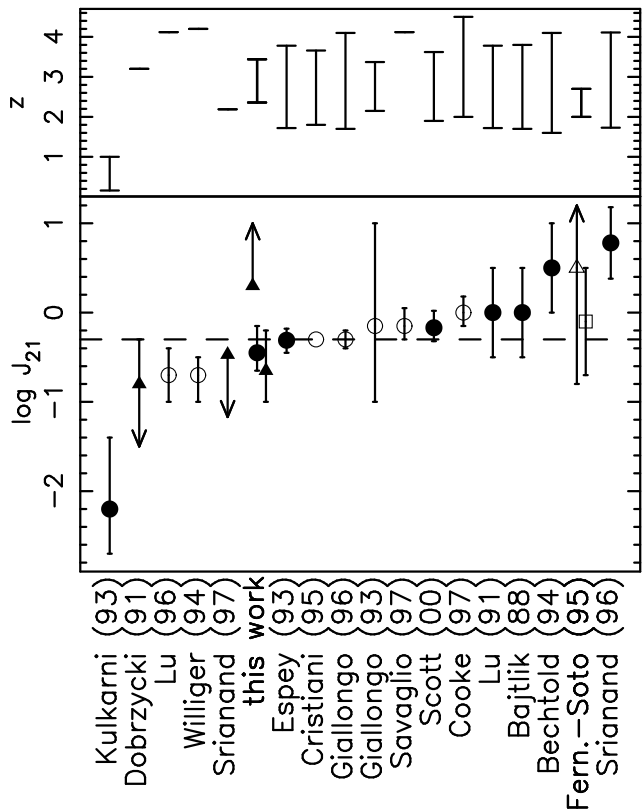}
\caption{Summary of previous proximity effect measurements of the mean
background Lyman limit intensity, $J$, in units of $10^{-21}$ ergs
s$^{-1}$ cm$^{-2}$ Hz$^{-1}$ sr$^{-1}$. Solid and open symbols
represent measurements from intermediate and high resolution data
respectively. Estimates from the background (foreground) proximity
effect are shown as circles (triangles), those that include both are
shown as squares. Error bars are $1\sigma$. The dashed line marks the
value of $J$ at $z = 2.5$--$3$ computed by Haardt \& Madau (1996) for
a background dominated by the observed QSO population and for a $q_0 =
0.1$ cosmology. The top panel shows the redshift intervals covered by
the various studies.}
\label{prev_j}
\end{figure}

We summarize these and other measurements of $J$ in Fig.\
\ref{prev_j}. It is worth noting that not all the points in this plot
are independent of one another. There are large overlaps in the data
used. For example, Q0014+813 is included in eight of these studies. In
addition, all except one of these measurements are based on `line
counting', i.e.\ the statistics of individual absorption lines. Only
\citeN{Lu96} considered the integrated absorption in $100\,$\AA\ bins.
\citeN{Zuo92} obtained a rough estimate of $J$ from $W^{-1}$
correlations (where $W$ is the rest equivalent width) and
\citeN{Moller92} used a basic flux statistics approach to investigate
the foreground proximity effect. Thus no alternative methods seem to
have been explored in great detail.

\nocite{Kulkarni93}
\nocite{Dobrzycki91}
\nocite{Lu96}
\nocite{Williger94}
\nocite{Srianand97}
\nocite{Espey93}
\nocite{Cristiani95b}
\nocite{Giallongo96}
\nocite{Giallongo93}
\nocite{Savaglio97}
\nocite{Scott00}
\nocite{Cooke97}
\nocite{Lu91}
\nocite{Bajtlik88}
\nocite{Bechtold94}
\nocite{Fernandez95}
\nocite{Srianand96}
\nocite{Haardt96}

Most authors have found good agreement between their data and the
ionization model of BDO and alternative explanations for the proximity
effect have not received much observational support. The sizes of
\lya\ absorbers inferred from observations of close QSO pairs (e.g.\
\citeNP{Dinshaw98} and references therein) seem to rule out the
possibility that the absorbers are too small to completely cover the
background QSO. In addition, \citeN{Lu91} found no difference between
the $W$ distribution of lines near QSOs and that of lines far from
QSOs. They also eliminated a broken power law for the redshift
distribution of lines as a possible cause for the proximity effect.

Thus increased ionization due to the extra UV flux from the QSO seems
to remain as the only credible explanation for the proximity
effect. However, it implies two observable effects:

1. The proximity effect should correlate with QSO luminosity. More
   luminous QSOs should deplete larger regions more thoroughly than
   less luminous ones. BDO claimed that this effect was present in
   their data and that it was consistent with the expectations from
   their ionization model. \citeN{Bechtold94} and \citeN{Scott00} also
   found a weak correlation. \citeN{Lu91} on the other hand found no
   evidence for a correlation at all but nevertheless concluded on the
   basis of simulations that this was consistent with the ionization
   model. \citeN{Srianand96} could not identify a correlation either.

2. In addition to the `classical' background proximity effect there
   should be a foreground proximity effect where the absorbing gas
   along the line of sight to a background QSO is depleted by the UV
   radiation of a close-by foreground QSO. Studying a triplet of QSOs
   separated by $2$ to $3$ arcmin \citeN{Crotts89} found no evidence
   for the existence of the foreground proximity effect.
   \citeN{Moller92} added another spectrum to this triplet and
   confirmed the negative result. \citeN{Dobrzycki91} observed a
   $\sim 10$~Mpc void in the spectrum of Q0302--003 with a foreground
   QSO separated from its line of sight by $17$ arcmin. However, the
   foreground QSO was displaced from the void by $\sim 3600$\kms,
   implying either that the QSO radiates anisotropically or that it
   turned on on a time-scale comparable to the light travel time from
   the QSO to the void. \citeN{Fernandez95} studied three QSO pairs
   separated by $3.8$ to $12.6$ arcmin but they were unable to reject
   the non-existence of the effect by more than $\sim 1 \, \sigma$.
   Finally, \citeN{Srianand97} reported a $\sim 7$~Mpc void in the
   spectrum of Tol 1038--2712 with a foreground QSO at the redshift of
   the void and separated by $17.9$ arcmin. Like \citeANP{Dobrzycki91}
   he showed that it was unlikely that the void was a chance
   occurrence. Thus there currently exists only a single example where
   an underdensity of absorption lines in the spectrum of a background
   QSO can be explained by the presence of a foreground QSO without
   making extra assumptions.

The main goals of this paper are to introduce a new method to analyse
QSO spectra for the proximity effect and to address the two problems
described above. The data used in this investigation are described in
Section~\ref{data}. They consist of the spectra of a close group of
$10$ QSOs which have not been included in any previous studies. Thus
the analysis presented here is independent from others in the sense
that it uses both a different method as well as different data. In
Section~\ref{cpe} we measure $J$ from the classical proximity effect
and demonstrate its correlation with QSO luminosity. In
Section~\ref{fpe} we turn to the foreground proximity effect. Finally,
we consider a range of uncertainties in Section~\ref{uncertainties}
and discuss our results in Section \ref{conclusions}.

Unless explicitly stated otherwise we use $q_0 = 0.5$, $\Lambda = 0$ and
$H_0 = 100 \, h$\kms\ Mpc$^{-1}$ throughout this paper.

\section{The data} \label{data}
\citeN{Williger96} performed a large \ion{C}{iv} survey in the South
Galactic Pole region. Here, we use that subset of the data which has
reasonable coverage of the \lya\ forest (10 spectra). The instrumental
resolution is $\sim 2\,$\AA\ and the signal-to-noise ratio per pixel
reaches up to 40 per $1\,$\AA\ pixel. The observations and the reduction
process are described in detail by \citeN{Williger96}.

\begin{figure}
\psfig{file=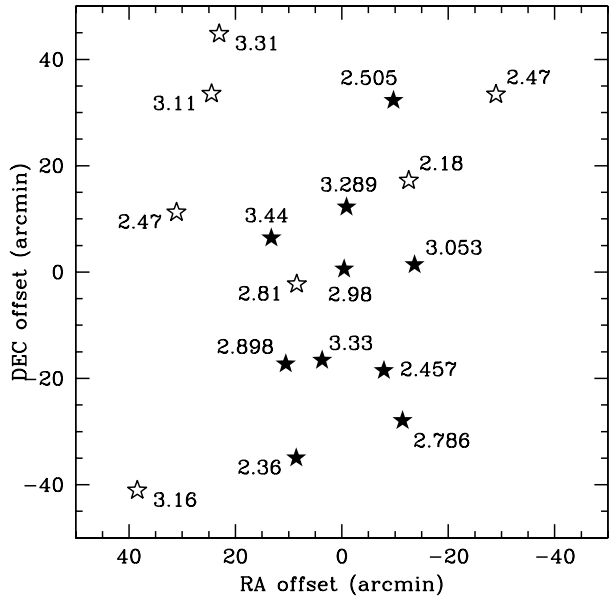,width=\columnwidth,silent=}
\caption{Distribution of QSOs in the sky. The field is centered on
$\alpha = 00^{{\rm h}} 42^{{\rm m}} 10^{{\rm s}}$ and $\delta =
-26\degr 40\arcmin$ (B1950). Solid stars mark the positions of the
QSOs whose spectra are analysed in this paper. Empty stars mark the
positions of additional foreground QSOs in the field. Emission
redshifts are indicated.}
\label{qsos}
\end{figure}

\begin{table}
\begin{minipage}{\columnwidth}
\caption{QSOs.}
\label{qsotab}
\begin{tabular}{crccrl}
\hline
Object & \multicolumn{1}{c}{$z_{\rm Q}$}
& $m_B^a$ & $\alpha$ & $L_{\nu}(912)^b$ & Refs. \\ 
\hline
Q0041--2607 &  $2.505$  &   $17.23$ &  $-0.60$ & $15.01$  & 1, 2 \\
Q0041--2638 &  $3.053$  &   $18.35$ &  $-0.53$ & $12.17$  & 1, 2 \\
Q0041--2658 &  $2.457$  &   $18.70$ &  $-0.88$ & $2.34 $  & 1, 2 \\
Q0041--2707 &  $2.786$  &   $18.03$ &          & $10.14$  & 1, 2 \\
Q0042--2627 &  $3.289$  &   $18.55$ &          & $13.29$  & 1, 2 \\
Q0042--2639 &  $2.98 $  &   $20.05$ &  $-0.09$ & $4.59 $  & 3 \\
Q0042--2656 &  $3.33 $  &   $19.55$ &  $-0.80$ & $4.11 $  & 3 \\
Q0042--2657 &  $2.898$  &   $18.78$ &          & $5.88 $  & 1, 2 \\
Q0042--2714 &  $2.36 $  &   $19.88$ &          & $1.09 $  & 4, 2, 5 \\
Q0043--2633 &  $3.44 $  &   $19.61$ &  $-0.77$ & $5.01 $  & 3 \\
\\
\multicolumn{6}{l}{Additional foreground QSOs:}\\
Q0040--2606 &  $2.47 $  &   $19.48$ &          & $1.80 $  & 4, 2 \\
Q0041--2622 &  $2.18 $  &   $19.28$ &          & $1.69 $  & 4, 2, 5 \\
Q0042--2642 &  $2.81 $  &   $20.60$ &  $-1.40$ & $0.28 $  & 3 \\
Q0043--2555 &  $3.31 $  &   $20.53$ &  $-0.89$ & $1.43 $  & 3 \\
Q0043--2606 &  $3.11 $  &   $20.37$ &  $-0.96$ & $1.07 $  & 3 \\
Q0044--2628 &  $2.47 $  &   $19.28$ &          & $2.18 $  & 4, 2, 5 \\
Q0044--2721 &  $3.16 $  &   $20.17$ &  $-0.45$ & $3.05 $  & 3 \\
\hline
\end{tabular}
$^a$The typical error on $m_B$ is $0.15$.\\
$^b$In units of $10^{30} \: h^{-2}$ ergs s$^{-1}$ Hz$^{-1}$.\\
References: (1) Redshifts and $B_J$ magnitudes from \citeN{Hewitt95}.
Where given, continuum slopes were measured from low-resolution
spectra kindly provided by Paul Francis.\\
(2) We converted from $B_J$ to Johnson $B$ using the colour equation
of \citeN{Blair82} and assuming $(B-V) = 0.3$.\\
(3) \citeN{Warren91b}. Conversion to $B$ magnitudes using the colour
equation of \citeN{Warren91}.\\
(4) \citeN{Drinkwater87}.\\
(5) Redshifts from \citeN{Williger96}.
\end{minipage}
\end{table}

A search of the literature revealed seven additional QSOs in this
field and in the appropriate redshift range. These will be considered
as potential foreground ionizing sources. The angular separations
range from $6.1$ to $95.8$~arcmin and the emission redshifts range from
$2.18$ to $3.44$. The distribution of all of these QSOs in the sky is
shown in Fig.~\ref{qsos}.

Since absolute spectrophotometry was not available for any of these
QSOs we had to estimate continuum flux densities from observed
$B$-band magnitudes. Assuming a power law continuum $f_\nu \propto
\nu^\alpha$ the observed flux density $f_\nu$ at observed wavelength
$\lambda$ is given by
\be \label{flux} 
f_\nu(\lambda) = \left[\frac{\lambda_X}{\lambda (1+z_{\rm Q})^{-1}}
\right]^\alpha (1+z_{\rm Q}) \: 10^{-0.4(m_X-k_X)} f_{\nu X}(0)
\ee 
where $\lambda_X$, $m_X$, $k_X$ and $f_{\nu X}(0)$ are the central
wavelength, observed magnitude, $K$-correction and 0-magnitude flux
\cite{Allen73} of the $X$-band respectively. For $X = V$ and $\alpha =
-0.6$ this equation gives a flux $1.6$ times higher than
\citeANP{Tytler87}'s \citeyear{Tytler87} empirical formula. However,
note that \citeANP{Tytler87} used the $K$-corrections of
\citeN{Evans77} whereas we use the $K$-corrections given by
\citeN{Cristiani90} because they extend beyond $z = 2.5$. For some
QSOs continuum slopes were not available. In these cases we used
$\alpha = -0.6$ which is similar to the value of $-0.5$ given by
\citeN{Francis93}.

We correct the above flux value for Galactic extinction by applying a
correction factor $10^{0.4 A(\lambda)}$ where
\be
A(\lambda) = R_V \: E(B-V) \: \frac{A(\lambda)}{A(V)}.
\ee
We use $R_V = A(V) / E(B-V) = 3.1$ which is the average value for the
diffuse interstellar medium \cite{Clayton88}. The variation of the
extinction with wavelength relative to that at $V$, $A(\lambda) /
A(V)$, is given by \citeN{ODonnell94} (optical) and \citeN{Cardelli89}
(UV). We take $E(B-V)$ for each QSO from the dust map of
\citeN{Schlegel98}. This procedure is equivalent to first correcting
$m_B$ for Galactic extinction and then applying a correction to
$\alpha$ due to the wavelength dependence of the extinction.

\citeN{Gaskell82} first pointed out that QSO redshifts measured from
high ionization emission lines like \lya\ or \ion{C}{iv} are
systematically lower than redshifts measured from lower ionization
lines like \ion{Mg}{ii} or the Balmer series which are thought to
indicate the systemic redshifts. When estimating $J$ from the
classical proximity effect using QSO redshifts derived from high
ionization lines, the result will be too high because the lower QSO
redshift implies a higher QSO flux at a given cloud and therefore (for
the same observed effect) a higher background. \citeN{Espey93} showed
that a velocity shift of $\sim1500$\kms\ lowered the value of
$J$ from $\log J = -20.75$ to $-21.30$ in \citeANP{Lu91}'s
\citeyear{Lu91} data.

The redshifts of the QSOs considered in this paper were all determined
from high ionization lines. However, since there is considerable
disagreement in the literature over the values of the line shifts and
possible correlations with QSO luminosity and/or emission line
properties (e.g. \citeNP{Tytler92} and references therein), it is
difficult to reliably correct for this effect. We shall therefore
resort to determining $J$ as a function of line shift in Section
\ref{j}.

In Table \ref{qsotab} we list all the QSOs considered as well as their 
redshifts, $B$-band magnitudes, continuum slopes and Lyman limit
luminosities calculated from equations \eref{flux} and \eref{lum}.

\section{Analysis} \label{analysis}
\subsection{Method}
Essentially, we use the method of \citeN{Liske98b} (hereafter LWC) to
look for regions of underdense absorption (i.e. `voids') in \lya\
forest spectra near the positions of suspected sources of ionizing
radiation. These sources may be the background QSOs themselves or
foreground sources near the line of sight to the background QSOs. The
technique does not rely on investigating variations of the number
density of individual absorption lines but rather uses the statistics
of the transmitted flux directly. Thus it elegantly sidesteps all
problems related to the incompleteness of lines due to limited
spectral resolution and signal-to-noise, $W$-limited versus
$N$-limited samples \cite{Chernomordik93,Srianand96} and Malmquist
bias \cite{Cooke97}. \citeN{Moller92} have previously employed a
somewhat more rudimentary version of our method to investigate
the foreground proximity effect. Flux statistics have also been used
in other contexts by \citeN{Liske00}, \citeN{Nusser00}, \citeN{Hui99},
\citeN{Croft99}, \citeN{Zuo94}, \citeN{ZuoLu93}, \citeN{Press93},
\citeN{PressR93}, \citeN{Webb92} and \citeN{Jenkins91}.

In order to identify large-scale regions of underdense absorption we
convolve a spectrum with a smoothing function in order to filter out
the high-frequency `noise' of individual absorption lines. A
large-scale feature will be enhanced most by this convolution if the
scale of the smoothing function, $\sigma_{\rm s}$, is similar to the
scale of the feature. Since we do not know the scale of any possible
features a priori we simply smooth the spectrum on all possible
scales, from 1 pixel where the data remain unchanged to the largest
possible scale where the data are compressed into a single number akin
to $1-D_A$ \cite{Oke82}. When plotted in the wavelength--smoothing
scale plane this procedure results in the `transmission triangle' of
the spectrum which we denote by $G(\lambda, \sigma_{\rm s})$ for the
case of a Gaussian smoothing function and by $T(\lambda, \sigma_{\rm
s})$ for the case of a top-hat smoothing function (Section \ref{imp}).

We then compare the observed transmission to the expected mean
transmission calculated on the basis of the simple null-hypothesis
that any \lya\ forest spectrum can be described as a collection of
absorption lines whose parameters are uncorrelated. We also calculate
the expected variance of the transmission in order to assess the
statistical significance of any fluctuations of the observed
transmission around the expected mean.

\subsection{Basic absorption model} \label{absmodel}
LCW considered a (normalized) \lya\ forest spectrum as a stochastic
process where each point in the spectrum is a random variable,
e$^{-\tau}$, drawn from the transmission probability density function.
They calculated the mean and variance of this function on the basis of
the above null-hypothesis. Including instrumental effects in their
calculation they then used these results to derive the mean and
variance of $G$,
\be \label{mG}
\langle G \rangle (\lambda, \sigma_{\rm s}) = 
\exp \left[-B \left(\frac{\lambda}{\lambda_\alpha}\right)^{\gamma+1}\right]
\ee
and
\be \label{sigG}
\sigma_G^2(\lambda, \sigma_{\rm s}) = \frac{\sigma_n^2(\lambda)}
{2\sqrt{\pi} \; \sigma_{\rm s}/ps} + \frac{\sigma_{{\rm e}^{-\tau}}^2(\lambda)}
{\sqrt{2\frac{\sigma_{\rm s}^2 + \sigma_{\rm LSF}^2}{q^2(\lambda)} + 1}},
\ee
where
\be \label{sigmaetau}
\sigma_{{\rm e}^{-\tau}}^2 = \exp \left[-2^{\beta-1} B 
\left(\frac{\lambda}{\lambda_\alpha}\right)^{\gamma+1}\right] - \;
\exp \left[-2 B \left(\frac{\lambda}{\lambda_\alpha}\right)^{\gamma+1}\right].
\ee
$\sigma_n$ denotes the noise of a spectrum, $\sigma_{\rm LSF}$ the
width of the instrumental line spread function, $ps$ the pixel size in
\AA\ and $\lambda_\alpha = 1216\,$\AA. $q$ is the intrinsic width
of the auto-covariance function of a `perfect' spectrum (i.e. before
it passes through the instrument) which we determine from simulations
(LWC; \citeNP{Liske00}). $B$ contains the normalization and is
given by
\be \label{B}
B = \frac{A}{\lambda_\alpha} \int \!\! \int N^{-\beta} f(b) \: W(N, b) \: 
\d N \d b,
\ee 
where $W(N, b)$ is the rest equivalent width of an absorption line of
column density $N$ and Doppler parameter $b$.
$B$ will be measured directly from the data (Section \ref{imp}).
$A, \gamma, \beta,$ and $f(b)$ are all part of the observationally
determined distribution of absorption line parameters
\be \label{dNdzdNdb}
\frac{\d^3 {\cal N}}{\d z \, \d N \, \d b} = A \: (1 + z)^\gamma N^{-\beta} 
f(b).
\ee
We take $\beta$ from recent high resolution studies as $1.5$
(\citeNP{Hu95}; \citeNP{Lu96}; \citeNP{Kim97}; \citeNP{Kirkman97}).
For the present data set \citeN{Williger00} determined $\gamma = 2.0$
which we will adopt. We discuss the effects of uncertainties in
$\beta$ and $\gamma$ in Section~\ref{depmod}.

\subsection{Incorporating local ionizing sources} \label{ionmodel}
In order to measure $J$ it is necessary to extend our model of the
transmitted flux to incorporate local fluctuations of the ionizing
radiation caused by discrete sources. We adopt the simple ionization
model of BDO which basically consists of the assumption
that the absorbing gas is highly photo-ionized so that an absorber's
column density is inversely proportional to the incident ionizing
flux. Thus in regions of enhanced ionizing radiation we must modify
equation \eref{dNdzdNdb}:
\be \label{new_dN}
\frac{\d^2 {\cal N}}{\d z \, \d N} \propto (1 + z)^\gamma N^{-\beta}
[1 + \omega(z)]^{1-\beta},
\ee
where
\be \label{om}
\omega = \frac{F_\nu(\lambda_{\rm LL})}{4\pi J(z)}.
\ee
$F_\nu(\lambda_{\rm LL})$ is the flux from the ionizing source (IS)
received by the absorber at wavelength $\lambda_{\rm LL} = 912\,$\AA\
in the restframe of the absorber. The IS may be the background QSO
itself or it may be a different, foreground QSO. The validity of
equation \eref{new_dN} is subject to the limitation that the spectral
shape of the background $J$ below the Lyman limit is similar to that
of the IS. This will be the case if the QSOs are the dominant
contributors to the background, if the IGM is optically thin (see
\citeNP{Espey93} for a discussion of the optically thick case) and if
the emission from the IGM does not drastically alter the shape of the
background \cite{Haardt96}.

$F_\nu$ may be calculated from
\be
F_\nu(\lambda_{\rm LL}) = \frac{L_\nu\left(\frac{\lambda_{\rm LL}}
{1+z_{\rm IS}'}\right)}{4\pi \: r_{\rm L}^2(z_a, z_{\rm IS}')} 
\: (1+z_{\rm IS}').
\ee
$r_{\rm L}(z_a, z_{\rm IS}')$ denotes the luminosity distance between
the absorber and the IS which, in general, is a function of the
absorber redshift, $z_a$, and the redshift of the IS as seen by the
absorber, $z_{\rm IS}'$. Since we will consider the foreground proximity
effect, where the absorber does not lie along the line of sight to the
IS, $z_{\rm IS}'$ is in turn a function of $z_a, z_{\rm IS},$ and the
angle $\alpha$ by which the absorber and the IS are separated on the
sky. See \citeN{Liske00b} on how to calculate $z_{\rm IS}'$ and $r_{\rm
L}(z_a, z_{\rm IS}, \alpha)$. Note the bandwidth correction factor
$(1+z_{\rm IS}')$ which is usually ignored at this point. The intrinsic
luminosity of the IS, $L_\nu$, is related to the observed flux at the
observed wavelength $\lambda$ by
\be \label{lum}
L_\nu\left(\frac{\lambda}{1+z_{\rm IS}}\right) = f_\nu(\lambda) \: 
\frac{4\pi \: r_{\rm L}^2(z_{\rm IS})}{1+z_{\rm IS}},
\ee
where $r_{\rm L}(z_{\rm IS})$ is the luminosity distance from Earth to
the IS. Thus we have
\be \label{omega}
\omega(z_a, z_{\rm IS}, \alpha) = \frac{f_\nu\left(\lambda_{\rm LL}
\frac{1+z_{\rm IS}}{1+z_{\rm IS}'}\right)}{4\pi J(z_a)}
\frac{1+z_{\rm IS}'}{1+z_{\rm IS}} 
\left[\frac{r_{\rm L}(z_{\rm IS})}{r_{\rm L}(z_a, z_{\rm IS}, \alpha)}
\right]^2.
\ee

The above ionization model is easily incorporated into our
transmission model. Because of equation \eref{B} the modification 
\eref{new_dN} implies
\be \label{Bz}
B \rightarrow B(z_a) = B \: 
\left[1+\omega(z_a, z_{\rm IS}, \alpha)\right]^{1-\beta}.
\ee

It is unlikely that we will be able to constrain the redshift
evolution of the background with the present sample as previous
studies of similar size but larger redshift coverage have been unable
to do so \cite{Cooke97,Giallongo96}. However, the background is
expected to peak smoothly in the redshift range covered here
\cite{Haardt96} and the near constancy of $J$ at $z > 2$ has been
supported by the consistency of simulations with the redshift
evolution of the \lya\ forest \cite{Dave99}. For these reasons we
assume $J(z) = {\rm const}$.

\subsection{Further modifications to the absorption model} \label{imp}
The inclusion of the proximity effect in our model renders one of the
approximations of LWC in the derivation of equations \eref{mG}
and \eref{sigG} invalid. There we used the fact that the mean (and the
variance) of the transmitted flux, $\langle{\rm e}^{-\tau}\rangle =
\exp[-B (\lambda/\lambda_\alpha)^{\gamma+1}]$, is approximately linear
in $\lambda$ over the scales of interest, so that, e.g.,
\bea
\langle G \rangle (\lambda, \sigma_{\rm s}) & = &
\frac{1}{\sqrt{2\pi} \: \sigma_{\rm s}} 
\int \langle{\rm e}^{-\tau}\rangle(\lambda')
\: \exp \left[-\frac{(\lambda - \lambda')^2}{2 \sigma_{\rm s}^2}\right]
\d\lambda' \nonumber \\
& \simeq & \exp \left[-B
\left(\frac{\lambda}{\lambda_\alpha}\right)^{\gamma+1}\right].
\eea
Because of the introduction of $\omega$, approximations like the one
above are no longer valid and we must now carry out all convolutions
explicitly:
\bea \label{newmG}
\lefteqn{\langle G \rangle_J (\lambda, \sigma_{\rm s}) =
\frac{1}{2\pi \: \sigma_{\rm LSF} \: \sigma_{\rm s}}
\int \!\!\! \int \exp \left[-B(z) \left(\frac{\lambda''}{\lambda_\alpha}\right)^{\gamma+1}\right]}
\nonumber \\
& & \times \: 
\exp \left[-\frac{(\lambda' - \lambda'')^2}{2 \sigma_{\rm LSF}^2}\right]
\: \exp \left[-\frac{(\lambda - \lambda')^2}{2 \sigma_{\rm s}^2}\right]
\d\lambda'' \d\lambda'
\eea
and
\bea \label{newsigG}
\lefteqn{\sigma_{GJ}^2(\lambda, \sigma_{\rm s}) = 
\frac{1}{\sqrt{2\pi} \: \hat \sigma_{\rm s}} \int \left[
\frac{\sigma_n^2(\lambda')}
{2\sqrt{\pi} \; \sigma_{\rm s}/ps} + 
\frac{\sigma_{{\rm e}^{-\tau}}^2(\lambda')}
{\sqrt{2\frac{\sigma_{\rm s}^2 + \sigma_{\rm LSF}^2}{q^2(\lambda')} + 1}}
\right]} \nonumber \\
& & \times
\exp \left[-\frac{(\lambda - \lambda')^2}{2 \hat \sigma_{\rm s}^2}\right]
\d\lambda',
\eea
where $\hat \sigma_{\rm s} = \sigma_{\rm s} / \sqrt{2}$.

When investigating the classical proximity effect it will be helpful
to use a top-hat smoothing function rather than a Gaussian because we
are working at the `edge' of the data. The equivalent of equations
\eref{newmG} and \eref{newsigG} are given by
\bea
\lefteqn{\langle T \rangle_J (\lambda, \sigma_{\rm s}) =
\frac{1}{\sqrt{2\pi} \: \sigma_{\rm LSF} \: 2\sigma_{\rm s}}
\int_{\lambda - \sigma_{\rm s}}^{\lambda + \sigma_{\rm s}} \!\!\! \int
\exp\left[-B(z)\left(\frac{\lambda''}{\lambda_\alpha}\right)^{\gamma+1}\right]}
\nonumber \\
& & \times \: 
\exp \left[-\frac{(\lambda' - \lambda'')^2}{2 \sigma_{\rm LSF}^2}\right]
\d\lambda'' \d\lambda'
\eea
and
\bea
\lefteqn{\sigma_{TJ}^2(\lambda, \sigma_{\rm s}) =
\frac{1}{2\sigma_{\rm s}}
\int_{\lambda - \sigma_{\rm s}}^{\lambda + \sigma_{\rm s}}
\left[ \frac{\sigma_n^2(\lambda')}{2\sigma_{\rm s} / ps} +
\frac{\sigma_{{\rm e}^{-\tau}}^2(\lambda')}
{\sqrt{2\frac{\sigma_{\rm LSF}^2}{q^2(\lambda')} + 1}} \right.}
\nonumber \\
& & \hspace{-3mm}\left. \times \frac{2}{2\sigma_{\rm s}} 
\int_0^{2\sigma_{\rm s} - 2 |\lambda - \lambda'|}
\hspace{-5mm}
\exp\left(-\frac{\lambda''^2}{2[2\sigma_{\rm LSF}^2 + q^2(\lambda')]}\right) 
\d\lambda''
\right] \d\lambda'.
\eea

As one approaches the background QSO in the classical proximity effect
the flux from the QSO increases and $B(z)$ decreases. Thus very close
to the QSO the model predicts a mean transmission of almost $1$ and a
variance of almost $0$. This is clearly unphysical as absorption lines
with $z_a \approx z_{\rm Q}$ and even with $z_a \ga z_{\rm Q}$ are frequently
observed. One of the reasons for this observation may be that
absorbers have peculiar velocities \cite{Srianand96,Loeb95}. We
accommodate peculiar velocities by convolving $B(z_a)$ with a Gaussian
of width $300$\kms.

We determined the normalization constant $B$ directly from the data.
First, we excluded all spectral regions with $\omega \ge 0.1$ from
background QSOs, assuming a fiducial value of $\log J = -21.0$.
For the present sample, the average size of the excluded regions
translates to $5000$\kms. For the remainder of each spectrum
we then computed $T(\sigma_{\rm s,max})$ where $\sigma_{\rm s,max}$
is the largest possible smoothing scale. To these we then fitted our
absorption model with $B$ as a free parameter.

\section{Results on classical proximity effect} \label{cpe}
\subsection{Significance} \label{cpesig}
With an absorption model and all its parameters in place we can
proceed by transforming the transmission triangle of a given spectrum, 
$T(\lambda, \sigma_{\rm s})$, to a `reduced transmission triangle' (RTT) 
by
\be
RT_J(\lambda, \sigma_{\rm s}) = \frac{T - \langle T \rangle_J}
{\sigma_{TJ}}.
\ee
The reduced triangle has the mean redshift evolution of the absorption
removed and shows the residual fluctuations of the \lya\ transmission
around its mean in terms of their statistical significance. When
neglecting the proximity effect in the calculation of $\langle T
\rangle$ (i.e.\ $J = \infty$), its presence in the data should be
revealed by a region of $RT_\infty > 0$ near the red edge of the
triangle.

We can consider the entire dataset at once in a compact manner by
constructing a combined RTT: we first shift the spectra into the
restframes of the QSOs, construct their transmission triangles and
then average them where they overlap. Since different lines of sight
are uncorrelated the variance of this composite is essentially just
$\sigma_T^2/n$, where $n$ is the number of spectra used. Thus at
rest wavelength $\lambda_r$ and at restframe smoothing scale
$\sigma_{{\rm s}r}$ we have
\bea \label{RT}
\lefteqn{
RT_\infty(\lambda_r, \sigma_{{\rm s}r}) =
\frac{1}{n} \sum \limits_{i = 1}^n \left\{ T_i[(1+z_{{\rm Q}i})\lambda_r, 
(1+z_{{\rm Q}i})\sigma_{{\rm s}r}] \right.} \nonumber \\
& & \left. \mbox{} - \langle T \rangle_{J = \infty} 
[(1+z_{{\rm Q}i})\lambda_r, 
(1+z_{{\rm Q}i})\sigma_{{\rm s}r}] \right\} \nonumber \\
& &  \times \left[ \frac{1}{n^2} \sum \limits_{i = 1}^n 
\sigma_{T J = \infty}^2
[(1+z_{{\rm Q}i})\lambda_r, (1+z_{{\rm Q}i})\sigma_{{\rm s}r}]
\right]^{-\frac{1}{2}},
\eea
where $T_i$ and $z_{{\rm Q}i}$ are the measured transmission and
redshift of the $i$th QSO and we use $J = \infty$ in the calculation
of $\langle T \rangle$ and $\sigma_T$.

The most significant `void' in this combined RTT lies at $(\lambda_r,
\sigma_{{\rm s}r}) = (1208.6\,$\AA, $3175$\kms), two pixels
away from the red edge of the triangle where one would expect a
signature from the proximity effect. It is significant at a level of
$5.2\sigma_T$. Excluding each of the individual spectra from the
composite in turn results in the significance of the feature varying
from $4.2\sigma_T$ to $6.2\sigma_T$, so that the effect is not
dominated by any single spectrum although there seems to be some
variation in the strength of the effect among the individual spectra.
This will be investigated in more detail in Section \ref{lumzcor}.

In this context it is helpful to ask what sort of signal one would
expect from the data if the ionization model of Section
\ref{ionmodel} were correct. In the next section we will answer this
question in detail with the help of simulations but one can already
gain a useful estimate by simply maximising the expectation value of
equation \eref{RT},
\be \label{m_RT}
\langle RT_\infty \rangle(J) =
\frac{\frac{1}{n} \sum_{i = 1}^n \left[ \langle T \rangle_J
- \langle T \rangle_{J = \infty} \right]}
{\left[ \frac{1}{n^2} \sum_{i = 1}^n \sigma_{T J = \infty}^2 \right]
^{\frac{1}{2}}},
\ee
with respect to $(\lambda_r, \sigma_{{\rm s}r})$. For example, for
$J_{21} = 1$ (where $J_{21} = J \times 10^{21}$ ergs s$^{-1}$
cm$^{-2}$ Hz$^{-1}$ sr$^{-1}$) we find a maximum of $5.0\sigma_T$.
Note that the expected significance of the signal is not so much a
function of the data quality but rather of the number of spectra
included in the analysis, since the variance of the transmission is
dominated by the `noise' of individual absorption lines.

An alternative explanation for the observed `void' would be a
systematic overestimation of the QSO redshifts. In this case we would
expect to see an effect similar to the one observed because we would
be including parts of the spectra in our analysis which correspond to
regions physically behind the QSOs and thus would show much less
absorption than expected. Note, however, that the redshifts of the
QSOs were determined from high ionization lines and are thus expected
to be too low, as discussed in Section \ref{data}, and not too high.
 
We conclude that we have detected a proximity effect at a significance
level of $> 99$ per cent.

\subsection{Measurement of $J$} \label{j}
The seemingly most straightforward way to derive an estimate of $J$
would be to directly fit our absorption model to the observed
composite transmission triangle with $J$ as a free parameter. Fitting
the transmission triangle rather than just the spectra has the
advantage of ensuring that the model fits {\em on all scales}. In the
previous section we saw that the data deviate most significantly from
a no proximity effect model at a scale of $\sim3000$\kms. On
the smallest smoothing scale the strongest deviation is only
$3\sigma_T$. Thus we can anticipate that by considering all scales we
may derive tighter constraints on $J$. However, the pixels of a
transmission triangle are strongly correlated with one another and
thus one would need to specify the entire covariance matrix in order
to judge the quality of a fit. Instead we will use a much simpler yet
effective approach which consists of considering only the most
significant positive and negative deviations of the model from the
data as a function of $J$.

We implement this approach by searching for the most significant local
extrema of an RTT in the region most likely to be affected by the
proximity effect. A local maximum (minimum), LMAX (LMIN), is defined
as any pixel in the RTT with $RT_J > 0$ ($RT_J < 0$) and where all
adjacent pixels have smaller (larger) values. If there is more than
one LMAX (LMIN) in a given wavelength bin (but at different smoothing
scales) only the most significant one is considered.

We then define the search region as all those pixels in the RTT for
which $\langle RT_\infty \rangle(J)$ is larger than some threshold
value $\eta$. We begin by setting $\eta = 1.0\sigma_T$ and search
for the most significant LMAX and LMIN in this region. If none are
found we increase the size of the search region by decreasing $\eta$
until the first LMAX (LMIN) has been found. Note that in order to
calculate $\langle RT_\infty \rangle(J)$ we need to assume some value
for $J$, which is precisely what we are trying to measure. However,
starting with $J_{21} = 1$ the procedure converges after only two
iterations. Even without iterating, the above procedure ensures that
the result does not depend sensitively on the exact value of $J$
(within sensible limits) chosen to define the search region.

\begin{figure}
\psfig{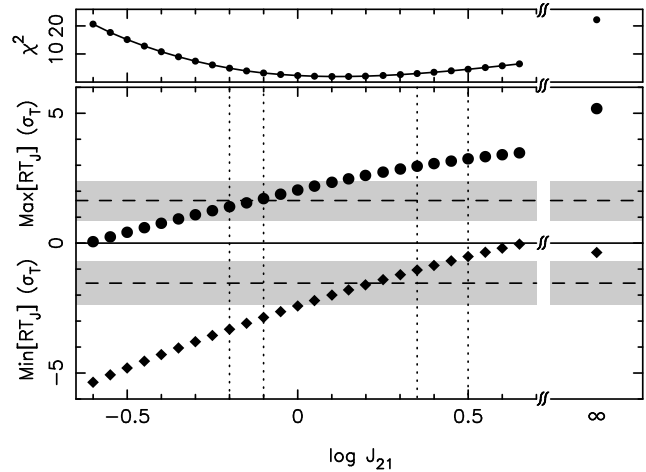}
\caption{Main panel: most significant positive (dots) and negative
(diamonds) deviations (in units of $\sigma_T$) of the absorption model
from the data in the region most likely affected by the proximity of
the QSOs as a function of the mean Lyman limit background intensity
$J$. The dashed lines and the grey regions mark the expected and $\pm
1 \sigma$ significance levels of these deviations in the case where
they are due to random fluctuations (i.e.\ for the correct model),
which were determined from $1000$ simulated datasets. The absence of a
proximity effect corresponds to $J = \infty$ and is strongly
rejected. Top panel: $\chi^2$ of the corresponding points in the main
panel. The best fit is achieved for $\log J_{21} = 0.1$ with
$\chi_{\rm min}^2 = 1.98$. The inner and outer dotted lines in the
main panel mark the formal $68$ and $90$ per cent confidence intervals
respectively.}
\label{j_bg}
\end{figure}

In Fig.~\ref{j_bg} (main panel) we plot both the most significant
positive and negative deviations of the model from the data detected
in this way as a function of the value of $J$ that was used in the
construction of the RTT. The void discussed in the previous section is
represented by the dot at $J = \infty$ (model with no proximity
effect). For models with $\log J_{21} > 0.0$ the RTT of the data
shows significant underdense absorption (i.e.\ maxima at $> 2
\sigma_T$). On the other hand, for small values of $J$ the model
predicts too little absorption on smaller scales and the RTT of the
data shows significant overdense absorption (i.e.\ significant minima)
for models with $\log J_{21} < 0.1$. Thus there seems to be no
value of $J$ for which the model is entirely consistent with the data.

However, recall that we are considering the {\em maximally} deviant
points. For the correct model, the expectation value of the difference
between a randomly chosen data point and the model is $0$. However,
given the additional information of the data point's rank, this is no
longer true. For example, the probability distribution function of the
maximum of a set of $n$ uncorrelated, normally distributed numbers is
given by $P(x) = n \, G(x) \, [\int_{-\infty}^x G(x') \d x']^{n-1}$, where
$G$ is the unit Gaussian. For $n >1$ both the mode and mean of $P(x)$
are $> 0$.

For an RTT the situation is more complicated, primarily because the
numbers from which the extrema are chosen are correlated. In order to
calculate an expectation value for the significance level of the
extrema one would thus have to specify the covariance matrix which is
exactly what we wanted to avoid. However, the expectation values of
the extrema and their correlation are easily obtained from
simulations.

We thus performed $1000$ simulations of the dataset ($= 10^4$ spectra,
hereafter S1) by randomly placing absorption lines according to the
parameters of Section \ref{absmodel} and using a constant S/N of
20. No proximity effect was included in the simulations. For each
dataset we then constructed its RTT (using $J = \infty$) and found the
most significant deviations of the model from the data. The mean
significance levels of these $1000$ maxima and minima are shown as the
dashed lines in Fig.~\ref{j_bg} and the $\pm 1 \sigma$ regions are
shown in grey.

We determine the best fit value of $J$ by calculating $\chi^2$ in the
top panel of Fig.~\ref{j_bg} for each of the data points in the main
panel. The best fit is achieved for $\log J_{21} =
0.1^{+0.4}_{-0.3}$, where the errors are the formal 90 per cent
confidence limits (outer dotted lines in Fig.~\ref{j_bg}). With
$\chi_{\rm min}^2 = 1.98$ the fit is acceptable ($P(\ge \chi_{\rm
min}^2) = 0.16$, where $P$ is the $\chi^2$-distribution with one
degree of freedom).

In order to check the error bars, the quality of the fit and the
validity of the procedure as a whole, we performed a second set of
$1000$ simulations (S2). This time we included the proximity effect
according to the ionization model of Section \ref{ionmodel} with $\log
J_{21} = 0.1$. Each of these datasets was analysed in the same
manner as the real data, i.e.\ for each we constructed Fig.~\ref{j_bg}
and measured $\log J$. On average, the presence of the proximity
effect is detected at the $5.2\sigma_T$ level. The mean of the
$1000$ $\log J$ measurements is $0.10$, $65$ per cent of the values
lie within the range $-0.15 < \log J_{21} < 0.35$ and $86$ per
cent of the values lie within the range $-0.3 < \log J_{21} <
0.5$. Finally, the fraction of measurements with $\chi^2_{\rm min} \ge
1.98$ is $0.10$. We thus conclude that our method works well and that
the error and quality of fit estimates above are reliable.

\subsubsection{Dependence on model parameters} \label{depmod}
Does the above result depend sensitively on any of the model
parameters introduced in Sections \ref{absmodel} and \ref{ionmodel}?
Because of the transition \eref{Bz} we must expect the result to
depend on $\beta$. At a given $J$, a larger value of $\beta$ increases
the model transmission and thus decreases both the maxima and minima
of Fig.~\ref{j_bg}, which will result in a larger measured value for
$J$. For $\beta = 1.7$ and $1.3$ we find $\log J_{21} =
0.6^{+0.3}_{-0.25}$ and $-0.45^{+0.45}_{-0.4}$ respectively. However,
increasing $\beta$ also has the effect of decreasing the model
variance so that the maxima and minima of Fig.~\ref{j_bg} move further
apart, which decreases the goodness of fit. For $\beta = 1.7$ we find
$\chi^2 = 10.1$ and thus the fit is no longer acceptable.

Since the redshift coverage of the present sample is not very large
and because we determine the optical depth normalization, $B$,
directly from the data, our results cannot depend sensitively on
$\gamma$. To confirm this we repeated the above analysis for $\gamma =
2.5$ and found $\log J_{21} = 0.15^{+0.35}_{-0.3}$.

Cosmological parameters enter the analysis via the last factor in
equation \eref{omega}. $h$ cancels out but the denominator is
relatively less sensitive to $q_0$ than the numerator. For an open
Universe this factor is larger than in the flat case and so a larger
$J$ will be measured. For $q_0 = 0.15$ we find $\log J_{21} =
0.15^{+0.4}_{-0.3}$ and thus our result does not depend sensitively on
the cosmological model.

\subsubsection{Emission line shifts}
\begin{figure}
\psfig{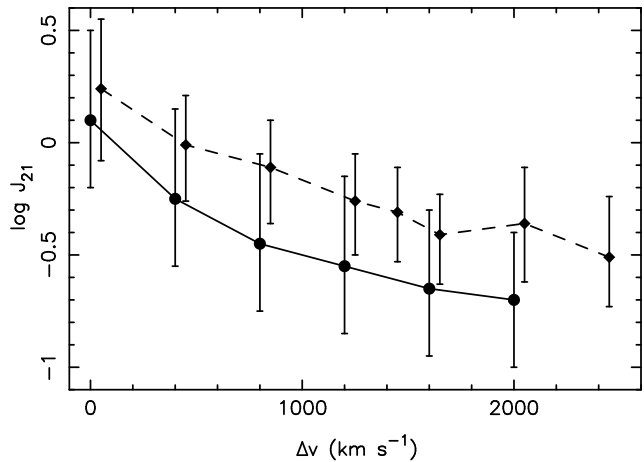}
\caption{Estimated value of the mean Lyman limit background intensity,
$J$, assuming that the measured high ionization emission line redshift
of every QSO is smaller than the true systemic value by an amount
$\Delta v$. Dots, solid line: this work. Diamonds, dashed line (offset
by $50$\kms\ for clarity): Espey's (1993) analysis of Lu
et~al.'s (1991) data for comparison. Error bars are $90$ per cent
confidence limits.}
\label{j_dv}
\end{figure}

In Section \ref{data} we already noted that the redshifts of the QSOs
have probably been underestimated since they were determined from high
ionization lines. Assuming that all the emission line redshifts are
offset from their true systemic values by a velocity $\Delta v$ to the
blue we have repeated the above analysis as a function of $\Delta
v$. In Fig.~\ref{j_dv} we show our estimate of $J$ for various values
of $\Delta v$ (dots, solid line). The reduction of $J$ is very similar
to that found by \citeN{Espey93} (diamonds, dashed line). From
similarly luminous QSOs \citeN{Espey93} estimated the mean velocity
shift for the \citeN{Lu91} QSOs to lie in the range $1300$\kms\
$< \Delta v < 1600$\kms\ and estimated the true background
intensity to be $\log J_{21} = -0.3^{+0.2}_{-0.22}$. For our
sample of QSOs the $\Delta v$-luminosity relationship given by
\citeN{Cooke97} predicts $\Delta v \approx 800$\kms\ which
yields $\log J_{21} = -0.45^{+0.4}_{-0.3}$. These results are in
good agreement with each other as well as with $\log J_{21}
\approx -0.3$ at $z = 2.5$--$3$ computed by \citeN{Haardt96} for a
background dominated by the QSO population observed in optical surveys
and a $q_0 = 0.1$ cosmology.

\nocite{Espey93}
\nocite{Lu91}

\subsection{Variation with luminosity and redshift} \label{lumzcor}
\subsubsection{Significance}
\begin{figure*}
\begin{minipage}{\textwidth}
\psfig{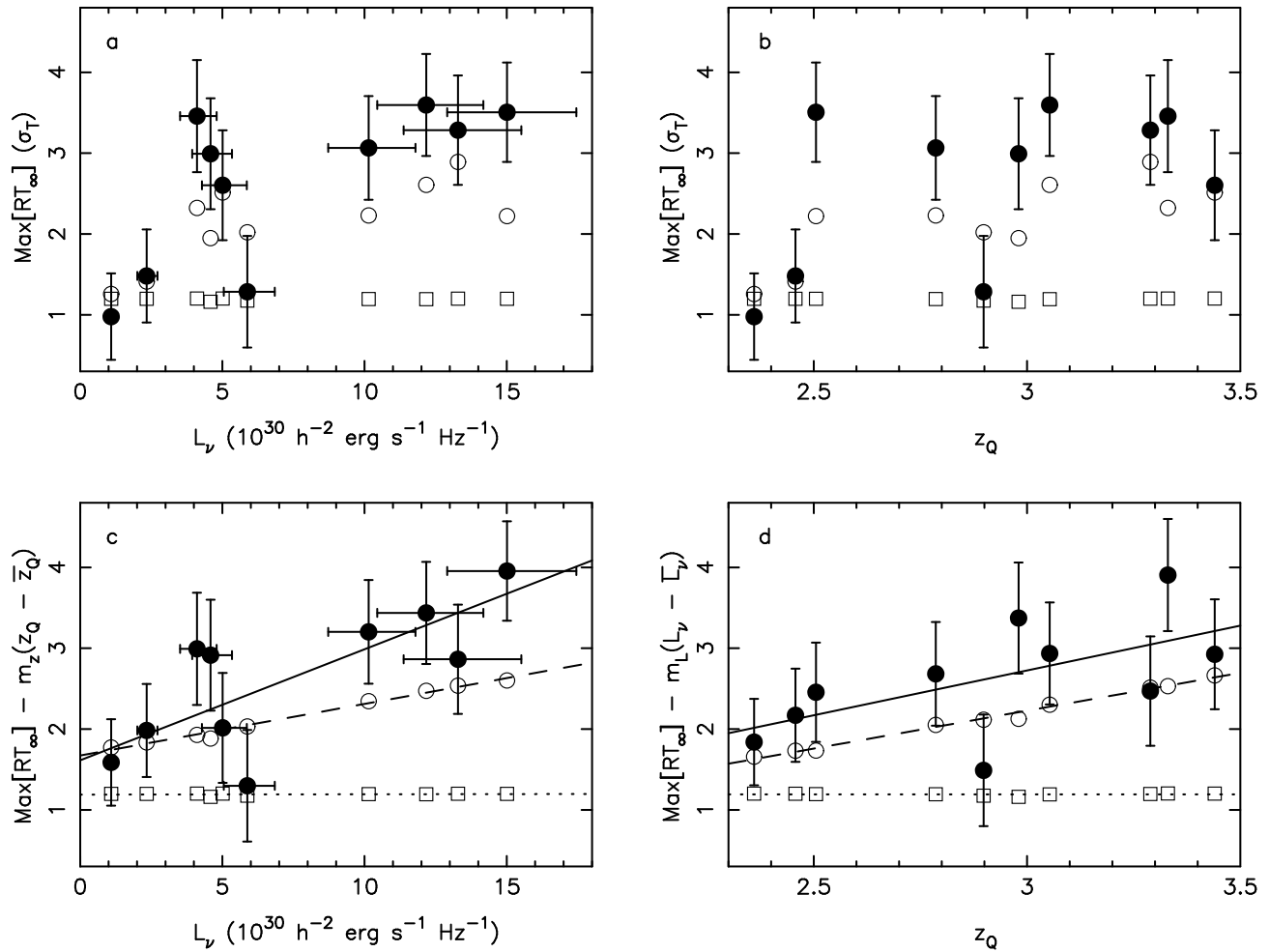}
\caption{(a) Significance of proximity effect in individual QSOs
versus QSO Lyman limit luminosity. Solid dots are the data. Open
circles are the mean significance levels found in $1000$ simulated
datasets, where the simulations include a proximity effect according
to the ionization model with $\log J_{21} = -0.45$ (S3). The
vertical error bars on the data are the $\pm 1\sigma$ ranges found in
these simulations. Open squares: same as open circles but with $J =
\infty$ (no proximity effect, S1). (b) Same as (a) but now plotted
against QSO redshift. (c) Same as (a) but all significance levels have
been scaled to the mean redshift of the sample using equation
\eref{fitfunc} in order to isolate the effect of luminosity on the
proximity effect. The solid, dashed and dotted lines show the best fit
$f(L_\nu, \bar z_{\rm Q})$ for the data and the two sets of simulations
respectively. (d) Same as (b) but all significance levels have been
scaled to the mean luminosity of the sample using equation
\eref{fitfunc} in order to isolate the effect of redshift on the
proximity effect. The solid, dashed and dotted lines show the best fit
$f(\bar L_\nu, z_{\rm Q})$ for the data and the two sets of simulations
respectively.}
\label{mlmz}
\end{minipage}
\end{figure*}

In Section \ref{cpesig} we noted that the significance of the
proximity effect varied somewhat when excluding individual spectra
from the combined RTT. We now examine the proximity effect in
individual spectra in order to test whether it is correlated with QSO
Lyman limit luminosity or redshift. The former correlation would be
expected if the proximity effect were due to the increased ionizing
flux in the vicinity of QSOs and the latter if in addition $J$ varied
with redshift and/or the luminosities and redshifts of the QSOs were
correlated.

We thus constructed the RTTs of the individual spectra (using
$J = \infty$) and searched for the most significant positive
deviations of the model from the data in the same way as described in
Section \ref{j}. In other words, for each QSO we determined the
significance of the proximity effect. In Fig.~\ref{mlmz}(a) we plot
these maxima as solid dots against QSO Lyman limit luminosity
(calculated from equation \eref{lum} and listed in Table
\ref{qsotab}). For comparison, we analysed two simulated datasets, S1
and S3, in exactly the same way. S1 was already introduced in Section
\ref{j} (no proximity effect model). S3 incorporates the proximity
effect according to the ionization model of Section \ref{ionmodel}
with $\log J_{21} = -0.45$ and assumes that all QSO redshifts have
been underestimated by $\Delta v = 800$\kms. Writing $M \equiv
{\rm Max}[RT_\infty]$, we plot as open circles the mean of $M$ found
in the $1000$ simulated datasets that include the proximity effect
(S3). The open squares are the same for the no proximity effect
simulations (S1).

Previous authors have presented similar plots (e.g.\ BDO's Fig.~1)
where they plotted the relative deficiency of absorption lines within
some constant radius of the QSOs. One of the advantages of our method
is that this radius is no longer constant but is rather allowed to
vary in order to maximise the significance of the missing absorption.

Let us first examine correlations in the simulated data. For S1 there is
clearly no correlation with luminosity which is as it should be. For
S3 there seems to be a trend of increasing significance with
increasing luminosity. However, the correlation does {\em not} seem to
be as tight as one might expect and there seems to be no well-defined
relation between the two. Considering that these points are the mean
of $1000$ simulations this can hardly be due to random error. The only
other parameter that varies from QSO to QSO in the simulations is the
redshift. In Fig.~\ref{mlmz}(b) we plot the significance levels
against QSO redshift. Again, for S1 there is no trend. S3 exhibits the
same sort of loose correlation as in Fig.~\ref{mlmz}(a). The
luminosities and redshifts of the QSOs are not significantly
correlated. Recall also that $J$ does not vary with redshift in our
simulations. Why then should there be a trend with redshift at all?
The reason is that a given underdensity of absorption is more
significantly detected when the `background' absorption line density
is higher than when it is lower (cf.\ LWC, Fig.~6b). Thus a given
QSO will have a more noticeable proximity effect at high redshift
(where the line density is higher) than at low redshift, all else
being equal.

We thus surmise that the lack of a well-defined relation between the
significance of the proximity effect, $M$, and luminosity for S3 in
Fig.~\ref{mlmz}(a) is due to the variation in redshift of the QSOs and
that $M$ is a function of both luminosity {\em and} redshift even
though the QSO luminosities and redshifts are not significantly
correlated and $J$ does not vary with redshift.

\begin{figure}
\psfig{file=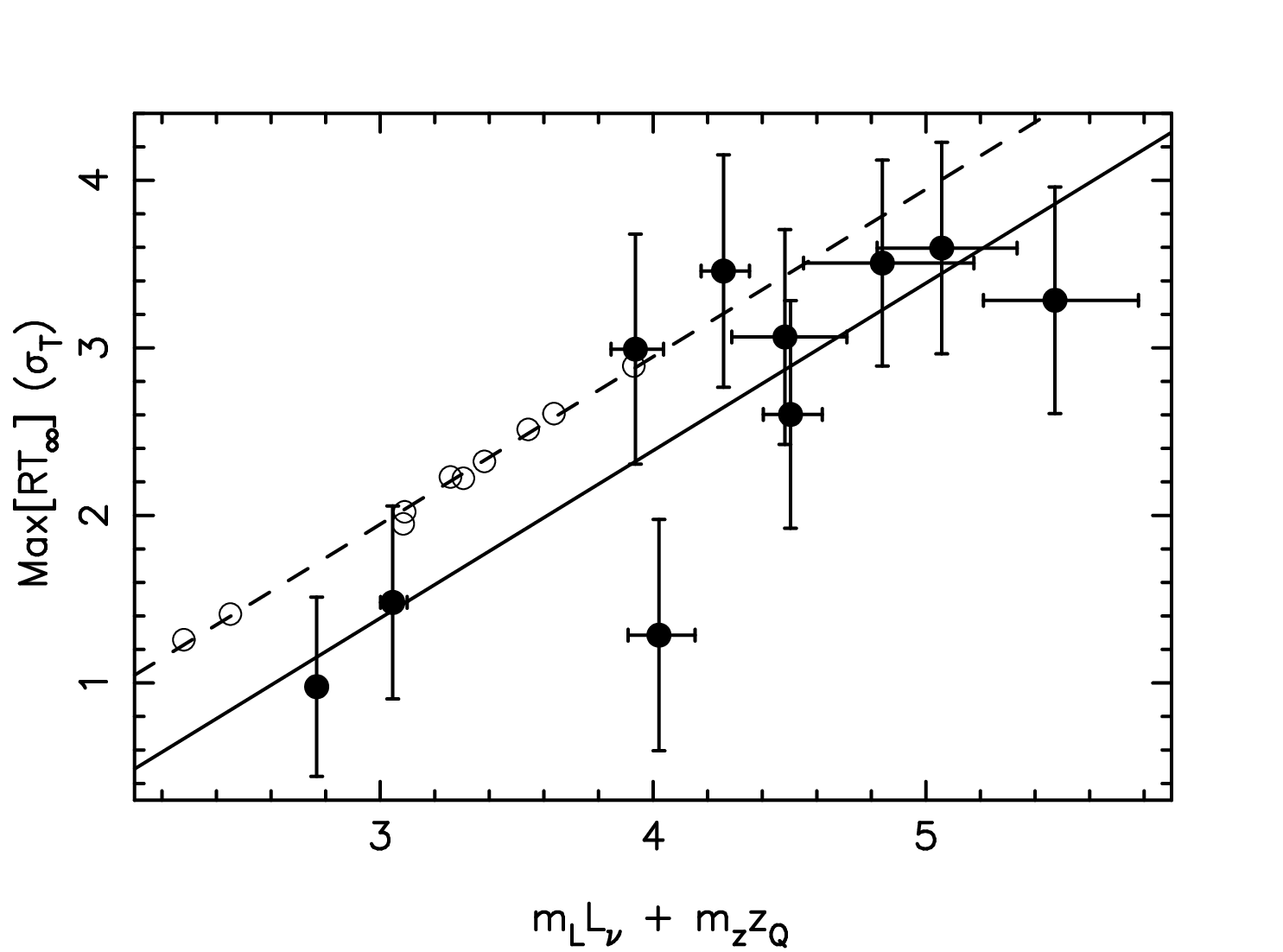,width=\columnwidth,angle=-90,silent=}
\caption{Significance of proximity effect in individual QSOs versus a
linear combination of QSO Lyman limit luminosity and redshift.
Symbols have the same meaning as in Fig.~\ref{mlmz}. The solid and
dashed lines show the best fit $f(L_\nu, z_{\rm Q})$ for the data and S3
respectively. S1 is not shown because for these points the best fit
slopes are $m_L \approx m_z \approx 0$.}
\label{mlz}
\end{figure}

To demonstrate this behaviour we first fit $M$ with the function
\be \label{fitfunc}
f(L_\nu, z_{\rm Q}) = c + m_L L_\nu + m_z z_{\rm Q}
\ee
and then scale the significance levels to the mean redshift, $\bar z$,
and mean luminosity, $\bar L_\nu$. The results are plotted in
Figs~\ref{mlmz}(c) and (d) respectively. In Fig.~\ref{mlz} we plot
$M$ against $m_L L_\nu + m_z z_{\rm Q}$. We can now see that S3
exhibits almost perfect correlation with both luminosity and redshift
and that the linear model \eref{fitfunc} gives a reasonably good
description of the simulations.

The discussion above implies that in order to properly disentangle
possible correlations of the proximity effect with luminosity and
redshift they should be determined jointly, not separately. We now
investigate this effect in more detail.
 
First, we need to choose a correlation statistic. Previous authors
have often used Spearman's rank correlation coefficient. However, in
our case it is reasonable to assume that $M$ is approximately Gaussian
distributed and this is indeed observed in the simulations. Therefore
it is not necessary to restrict ourselves to a non-parametric test. In
addition we have already shown that the linear model \eref{fitfunc}
gives a good description of the simulations and from Fig.~\ref{mlz} we
can judge that it is also an acceptable model for the real data (see
also Fig.~\ref{chisq}c). It is therefore reasonable to use the slopes
of a linear $\chi^2$ fit as correlation measures. This also has the
advantage that the arguments of the following paragraphs can be
understood analytically. In any case, we found that these arguments
are qualitatively reproduced when using Spearman's rank correlation
coefficient.

\begin{figure*}
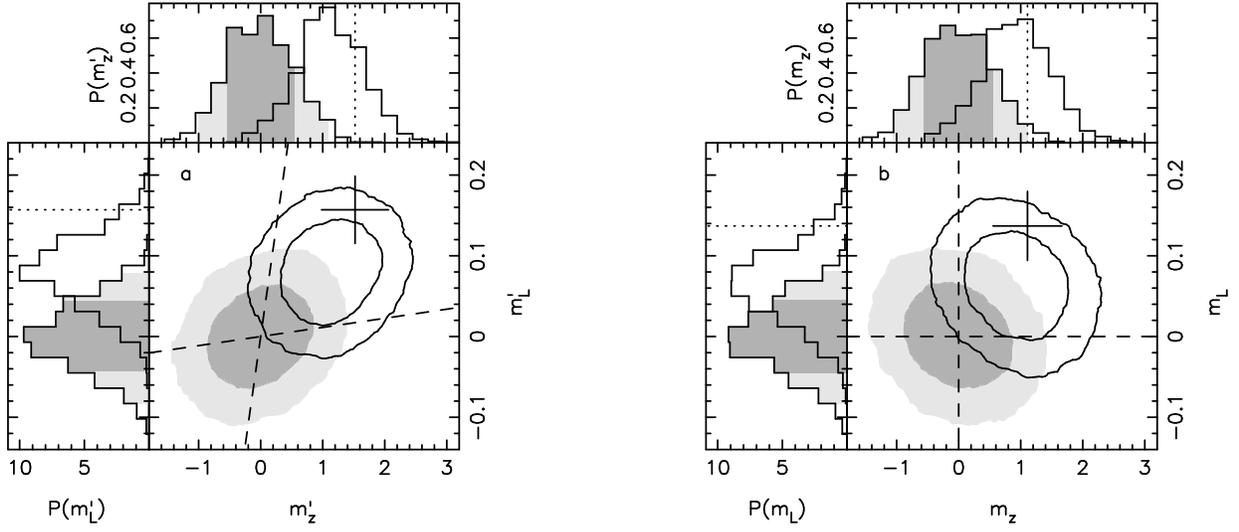

\begin{minipage}{\columnwidth}
\psfig{file=fig7a.ps,width=\textwidth,angle=-90,silent=}
\end{minipage}
\hfill
\begin{minipage}{\columnwidth}
\psfig{file=fig7b.ps,width=\textwidth,angle=-90,silent=}
\end{minipage}
\caption{(a) The dark and light grey shaded regions show the 68 and 95
per cent confidence regions of $(m_z', m_L')$ derived by independently
fitting the significance of the proximity effect in the $1000$
simulated datasets of S1 (no proximity effect) with \eref{fitfuncs}.
The contour lines are the same for S3 (ionization model). The cross
marks the measured values and formal $1\sigma$ errorbars of $(m_z',
m_L')$ for the real data. The dashed lines correspond to the lines
$m_z = 0$ and $m_L = 0$ in panel (b). The upper and left panels show
the one-dimensional probability distributions of $m_z'$ and $m_L'$
respectively. The dark and light shaded regions in these panels are
the one-dimensional 68 and 95 per cent confidence regions
respectively. (b) Same as (a) for $(m_z, m_L)$, derived by fitting the
significance of the proximity effect with \eref{fitfunc}.}
\label{pmlz}
\end{figure*}

The simplest thing we can now do is to determine the slopes for the
$M$-$L_\nu$ and $M$-$z_{\rm Q}$ relations independently by fitting the
functions
\be \label{fitfuncs}
\begin{array}{lcl}
f_L(L_\nu) & = & c_L + m_L' L_\nu \\
f_z(z_{\rm Q}) & = & c_z + m_z' z_{\rm Q}
\end{array}
\ee
to $M$. We have done this for the $1000$ simulated datasets of both S1
and S3 as well as for the real data. The results are shown in
Fig.~\ref{pmlz}(a). The shaded regions in the main panel are the $68$
and $95$ per cent confidence regions for S1, the contours are the same
for S3 and the cross marks the result for the real data.

The first thing we notice is that the values of $(m_z', m_L')$
measured in the real data are consistent with the ionization model,
but they are inconsistent with the no proximity effect model at $>
3\sigma$. However, note that in the simulations $m_L'$ and $m_z'$ are
not independent. Calculating Pearson's correlation coefficient we find
\be \rho(m_L', m_z') = \hat \rho(L_\nu, z_{\rm Q}) = 0.14, 
\ee 
where $\hat \rho(L_\nu, z_{\rm Q})$ is the correlation coefficient
between $L_\nu$ and $z_{\rm Q}$ for the $10$ QSOs used here, not that
of the parent population ($= \rho(L_\nu, z_{\rm Q})$) from which they
were drawn, which may well be zero. Thus any measurement in the top
right-hand part of the plot deviates from the no proximity effect
hypothesis less significantly than what would have been inferred if
the correlation between $L_\nu$ and $z_{\rm Q}$ had been
neglected. Note that it is the actual numerical value of $\hat \rho$
that matters here and not whether or not it is consistent with
$\rho(L_\nu, z_{\rm Q}) = 0$.

What would we expect if the proximity effect were caused by some
property, $x$, of the QSO or its environment unrelated to the QSO's UV
flux (`$x$-model')? As for the ionization model we would expect a
correlation of $M$ with $z_{\rm Q}$ because this is simply the effect
of increasing absorption line density with increasing redshift. Since
$\hat \rho(L_\nu, z_{\rm Q}) \neq 0$ we therefore expect the centre of
the shaded ellipses in Fig.~\ref{pmlz}(a) to move along the line
\be
m_L' = m^z_L \; m_z' = \hat \rho(L_\nu, z_{\rm Q}) \; 
\sqrt{\frac{V(z_{\rm Q})}{V(L_\nu)}} \; m_z'
\ee
(marked by a dashed lined), where $V$ denotes the sample variance.
Thus a measurement in the top right-hand part of the plot deviates
less significantly from the $x$-model than what would have been
inferred if the correlation between $L_\nu$ and $z_{\rm Q}$ had been
neglected.

This complication can be avoided if we use $(m_z, m_L)$ of equation
\eref{fitfunc} instead of $(m_z', m_L')$. Fig.~\ref{pmlz}(b) shows the
result of fitting \eref{fitfunc} to the simulations and the data. The
relationship between Figs~\ref{pmlz}(a) and (b) can be most easily
understood by writing down the linear co-ordinate transformation which
relates $(m_z', m_L')$ and $(m_z, m_L)$ as
\be
{m_z \choose m_L} = \frac{1}{1 - \hat \rho^2}
\left( \begin{array}{cc}
1 & -m^L_z \\
-m^z_L & 1 \\
\end{array} \right )
{m_z' \choose m_L'}.
\ee
We can now see that they are related by a Lorentz transformation
followed by a stretch of $(1 - \hat \rho^2)^{-\frac{1}{2}}$ and that
the line $m_L' = m^z_L \; m_z'$ is transformed to the line $m_L = 0$.
Thus in this frame of reference we do not have to worry about $\hat
\rho$ when asking whether the data is compatible with the $x$-model,
except for the fact that we now have $\rho(m_L, m_z) = -\hat
\rho(L_\nu, z_{\rm Q})$.

In Figs~\ref{chisq}(a), (b) and (c) we plot the distribution of
minimum $\chi^2$ values obtained from fitting \eref{fitfuncs} and
\eref{fitfunc} respectively. The shaded histogram (S1) follows the
$\chi^2$-distribution quite well in all cases. However, both the real
data (dotted line) and the ionization model (S3, solid histogram) are
not well modelled by the independent fits \eref{fitfuncs}.

Thus we conclude that the observed significance of the proximity
effect is a linear function of both redshift and luminosity and is
well described by a function of the form \eref{fitfunc}. This observed
correlation with luminosity {\em and} redshift is inconsistent with
the no proximity effect model at the $3.5\sigma$ level. A model that
exhibits correlation with redshift but not luminosity is excluded at
the $2.6\sigma$ level. These values take into account the correlation
between $L_\nu$ and $z_{\rm Q}$ of the present dataset. If we had
ignored this correlation we would have inferred $4.1\sigma$ and
$3.2\sigma$ respectively.

The discussion above enables us to go one step further. Consider again
the $x$-model. If $x$ is indeed uncorrelated with $L_\nu$ for the
general QSO population, i.e.\ $\rho(x, L_\nu) = 0$, then for our
sample we would most likely find $\hat \rho(x, L_\nu) = 0$, which we
have implicitly assumed in the previous discussion. However, $\hat
\rho(x, L_\nu)$ may well be $\neq 0$, either because of random
fluctuations or because $\rho(x, L_\nu) \neq 0$. This would induce a
spurious correlation between $M$ and $L_\nu$. What value of $\hat
\rho(x, L_\nu)$ is required so that the $x$-model is consistent with
the data? Assuming $\hat \rho(x, z_{\rm Q}) = 0$ we find that $\hat
\rho(x, L_\nu)$ has to lie in the range
\be
0.37 < \hat \rho(x, L_\nu) < 0.80
\ee
in order to be consistent with the data at the $1\sigma$ level. Thus
in the present dataset, the hypothetical property $x$ would have to be
noticeably correlated with luminosity. 

Since we do not know the distribution from which $x$ is drawn it is
not possible to reliably estimate whether this result excludes the
hypothesis that $x$ and $L_\nu$ are not correlated in the parent
population. However, assuming Gaussianity in both $x$ and $L_\nu$,
we find that the most likely value of $\hat \rho(x, L_\nu) = 0.59$
excludes the hypothesis $\rho(x, L_\nu) = 0$ at the $\sim 96$ per cent
confidence level (using Student's $t$-distribution).

By comparing the data with S3 in Fig.~\ref{pmlz}(b) it is apparent
that the observed dependence of the proximity effect on redshift is
entirely accounted for by the evolution of the number density of
absorption lines and there is no evidence to suggest that $J$ varies
over the redshift range considered here. It is also apparent that this
conclusion is not overly sensitive to our choice of $\gamma$, the
evolutionary index of the absorption line density. On the other hand,
the data do not exclude some (positive or negative) evolution either.
In any case, if $J$ is due to the known QSO population then it is
expected to peak smoothly in the redshift range covered here
\cite{Haardt96} and thus one would not expect strong evolution.

\begin{figure}
\psfig{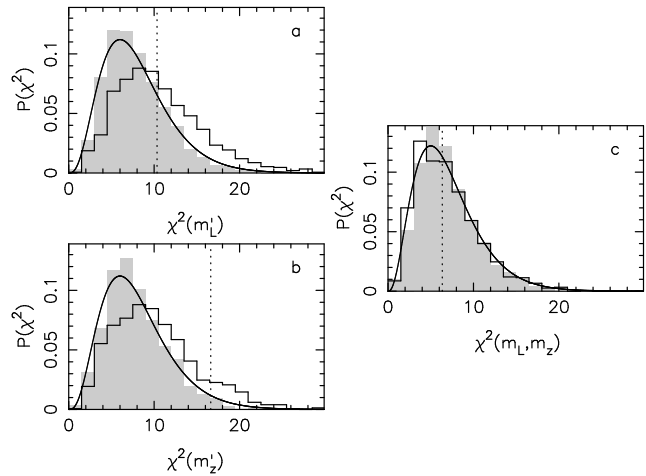}
\caption{(a) The grey shaded histogram is the distribution of minimum
$\chi^2$ values obtained from fitting $f_L(L_\nu)$ (equation
\ref{fitfuncs}) to the $1000$ simulated datasets of S1 (no proximity
effect). The solid histogram is the same for S3 (ionization
model). The dashed line marks the value of $\chi^2$ for the real
data. The smooth solid line shows the $\chi^2$-distribution for eight
degrees of freedom. (b) Same as (a) for $f_z(z_{\rm Q})$. (c) Same as
(a) and (b) for $f(L_\nu, z_{\rm Q})$ (equation \ref{fitfunc}). The
solid line shows the $\chi^2$-distribution for seven degrees of
freedom.}
\label{chisq}
\end{figure}

\subsubsection{Size}
If the ionization model is correct one would in principle also expect
the size of the region affected by the QSO's UV flux to be correlated
with luminosity. We can test for this effect by noting the size of the
smoothing function at which the maxima $M$ of Fig.~\ref{mlmz} were
detected. However, in the simulations we find that the distribution of
sizes is very broad and almost one-tailed and thus difficult to
characterise. In addition the distributions for S1 and S3 overlap
almost completely with only the peak moving to slightly larger values
for S3. There is also some evidence that for the ionization model the
distribution of sizes moves to larger values for larger luminosities
but again the effect is small compared to the extent of the
distribution. Thus we are forced to conclude that this is not a
powerful test and simply note that the detected sizes in the real data
range from $\sim 500$ to $\sim 7000$\kms\ and lie well within
the $90$ per cent confidence ranges of both S1 and S3 in all cases.

\section{Results on foreground proximity effect} \label{fpe}
\subsection{Existence of the effect}
We now exploit the fact that the QSOs of Table~\ref{qsotab} are a
close group in the plane of the sky. Essentially, we repeat here the
analysis of Section~\ref{cpesig}: for each pair of background QSO
(BQSO) and foreground ionizing source (IS), we shift the spectrum of
the BQSO into the restframe of the IS and construct its transmission
triangle using $J = \infty$. We now use a Gaussian smoothing function
because we are no longer working at the `edge' of the data. We then
average all the transmission triangles where they overlap, creating a
composite RTT. Thus in equation~\eref{RT} $T_i$ still refers to the
measured transmission of the BQSO (but is replaced by $G_i$ because we
now use a Gaussian smoothing function) and we replace $z_{{\rm Q}i}$
with $z_{{\rm IS}i}$, where $i$ now labels a BQSO-IS pair. If the
foreground proximity effect exists, it should be more significant in
this composite RTT than in any individual RTTs and is expected to
appear as a region of $RG_\infty > 0$ near $\lambda_r =
\lambda_\alpha$. From now on we exclude all spectral regions within
$5000$\kms\ of the BQSOs from the analysis in order not to
contaminate our results with the background proximity effect.

Before we proceed we need to consider the following complication. For
each BQSO-IS pair we can calculate $\omega_{\rm p} = \omega(z=z_{\rm
IS}, z_{\rm IS}, \alpha)$ (assuming $\log J_{21} = -0.45$, cf.\
equation~\ref{omega}). $\omega_{\rm p}$ is the maximum value a given
IS can achieve along the line of sight to a BQSO, separated on the sky
by an angle $\alpha$. Pairs with large $\omega_{\rm p}$ should show
a strong proximity effect. However, below some value of $\omega_{\rm
p}$ the proximity effect will be essentially non-existent. Adding
pairs with $\omega_{\rm p}$ below this value to the composite may in
fact decrease the overall significance of the effect.

We thus construct several composite RTTs, each with a different lower
limit on $\omega_{\rm p}$, which we denote by $\omega_{\rm c}$. Thus
$RG_J(\omega_{\rm c})$ refers to a composite reduced transmission
triangle in which all BQSO-IS pairs with $\omega_{\rm p} > \omega_{\rm
c}$ are included.

\begin{figure}
\psfig{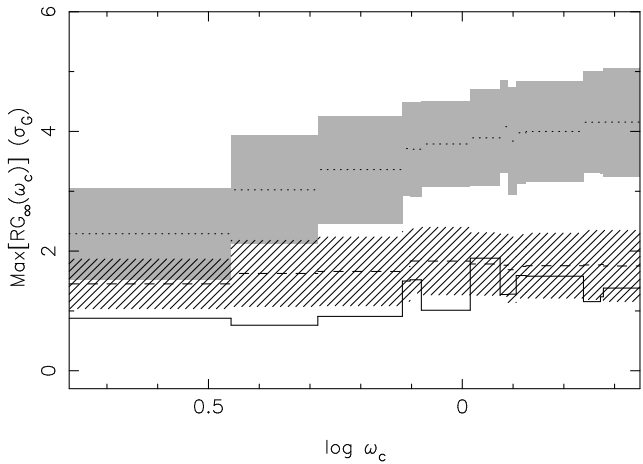}
\caption{Significance of the foreground proximity effect as derived
from composite RTTs which include all BQSO-IS pairs with $\omega_{\rm
p} > \omega_{\rm c}$. The solid line is the data. The dotted line and
grey shaded regions are the mean and $\pm 1\sigma$ significance levels
found in $1000$ simulated datasets that include the foreground
proximity effect according to the ionization model with $\log J_{21} =
-0.45$ (S4). The dashed line and hashed region are the same for $J =
\infty$ (no proximity effect, S1). The number of BQSO-IS pairs
included in the RTT increases from $1$ at the left end of the plot to
$14$ at the right end.}
\label{mwp}
\end{figure}

Each of these RTTs was searched for the most significant positive
deviation of the model from the data in the interval
[$-1000$\kms$, +2000$\kms] around $\lambda_\alpha$. The
interval is asymmetric to allow for underestimated QSO redshifts. In
Fig.~\ref{mwp} we plot these maxima versus $\omega_c$. For comparison
we have performed the same analysis on the $1000$ simulated datasets
of S1 (no proximity effect). We plot the mean and $\pm1\sigma$
significance levels as the dashed line and hashed region
respectively. Clearly, our data are consistent with the absence of any
foreground proximity effect.

Given the luminosities and inter-sightline spacings of the present set
of QSOs, do we expect to be able to detect a signal? To answer this
question we have created a fourth set of $1000$ simulations
(S4). These include the effects of all foreground IS with $\omega_{\rm
p} > 0.5$ and $\log J_{21} = -0.45$. Subjecting S4 to the same
analysis as above yields the dotted line and grey shaded region in
Fig.~\ref{mwp}. Evidently, if $J$ has indeed the value that was
measured from the background proximity effect and if QSOs radiate
isotropically, then we should be able to detect the foreground
proximity effect at the $2-4\sigma$ level in our data.

\subsection{Lower limit on $J$}
\begin{figure}
\psfig{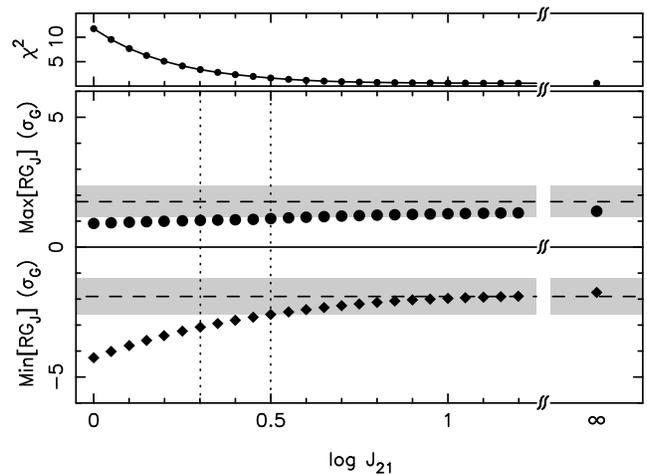}
\caption{This plot is the equivalent of Fig.~\ref{j_bg} for the
foreground proximity effect. In the main panel we show the most
significant positive (dots) and negative (diamonds) deviations (in
units of $\sigma_G$) of the absorption model from the data in the
regions near the positions of close-by foreground QSOs as a function
of the mean Lyman limit background intensity $J$. The dashed lines and
the grey regions mark the expected and $\pm 1 \sigma$ significance
levels of these deviations in the case where they are due to random
fluctuations. The data are consistent with the absence of a foreground
proximity effect ($J = \infty$). Thus we cannot derive an upper limit
on $J$. The two dotted lines mark the $68$ and $90$ per cent confidenc
lower limits on $J$.}
\label{j_fg}
\end{figure}

The above result implies that we can at least derive a lower limit to
$J$ under the assumption that QSOs radiate isotropically. Setting
$\omega_{\rm c} = 0.5$ we have repeated the analysis of
Section~\ref{j} using a foreground RTT that includes $14$ BQSO-IS
pairs. The result is shown in Fig.~\ref{j_fg}. We can see that the
data are consistent with a large range of values of $J$, including $J
= \infty$. However, for small values of $J$ the model predicts too
little absorption to be compatible with the data. From this constraint
we derive a lower limit of $\log J_{21} > 0.3$ ($90$ per cent
confidence).

From Fig.~\ref{j_dv} we can see that this lower limit is larger than
the upper limit derived from the background proximity effect for all
$\Delta v \ga 200$\kms. If we did not underestimate $J$ in
Section~\ref{j} then the simplest explanation for this discrepancy is
that QSOs radiate anisotropically. If we require that the lower limit
derived from the lack of a foreground proximity effect coincides with
the upper limit of the background measurement then the QSOs of our
sample must emit less ionizing radiation in the plane of the sky than
along the line of sight to Earth by at least a factor of $1.4$ for
$\Delta v > 400$\kms. This number increases to $2.2$ for $\Delta v >
800$\kms.

\subsection{The Q0042--2639 quadrangle}
Having established that the dataset as a whole does not exhibit any
evidence for the existence of the foreground proximity effect, we now
present a possible exception to this rule. From Fig.~\ref{qsos} we can
see that Q0042--2639 is surrounded by four nearby background
QSOs. Interestingly, all four show underdense absorption near the
position of the foreground QSO. In Table~\ref{quadtab} we list the
BQSO name, the angular separation from the foreground QSO, the
significance of the underdensity, $M$ (in units of $\sigma_G$), the
velocity offset of the underdensity from the foreground QSO redshift
and finally the size of the underdensity (FWHM of the smoothing
Gaussian). In the last line we list the same quantities for the
composite RTT of the four BQSOs. Note that the offsets from the
foreground QSO's redshift are of the same magnitude but of opposite
sign for the two pairs of opposing BQSOs (cf.\ Fig.~\ref{qsos}).
However, this is no longer true if we add $\ga 400$\kms\ to the
foreground QSO's redshift.

At first glance, it may seem exceedingly unlikely to find an
underdensity in four different lines of sight at a similar position
(which happens to coincide with a foreground QSO) by chance. However,
the spread of the underdensities in redshift is actually fairly
large. Thus when we combine the four lines of sight to a composite RTT
the significance level of the `void' rises only marginally from $\sim
3\sigma_G$ in individual lines of sight to $3.5\sigma_G$ in the
composite. Nevertheless, in the $1000$ simulated datasets of S1 we
find that the largest random fluctuations in the composite RTT have a
mean significance level of $(1.69 \pm 0.53) \sigma_G$ and thus it
seems that the void is in fact genuine. Indeed, a measurement of $J$
from the composite RTT yields $\log J_{21} = -0.65^{+0.6}_{-0.5}$
which is lower than, but consistent with our measurements from the
background proximity effect. From Fig.~\ref{mlmz} (and
Table~\ref{qsotab}) we can see that Q0042--2639 also shows a slightly
(but not significantly) more prominent background proximity effect than
`predicted' by the other QSOs for its luminosity and redshift.
Together these observations may be an indication that the luminosity
of Q0042--2639 has been underestimated.

\begin{table}
\caption{The Q0042--2639 quadrangle.}
\label{quadtab}
\begin{tabular}{lccrc}
\hline 
BQSO & $\alpha$ & $M$ & \multicolumn{1}{c}{$\Delta$} & FWHM$_{\rm s}$\\
& arcmin & $\sigma_G$ & \kms & \kms \\
\hline
Q0043--2633 & $13.6$ & $2.97$ & $-460$ & $1260$ \\
Q0042--2627 & $11.7$ & $3.11$ & $-2390$ & $5530$ \\
Q0041--2638 & $11.9$ & $2.71$ & $+230$ & $3120$ \\
Q0042--2656 & $17.6$ & $3.03$ & $+2630$ & $1210$ \\
Composite & & $3.52$ & $+160$ & $1880$ \\
\hline
\end{tabular}
\end{table}

\section{Uncertainties} \label{uncertainties}
In Section~\ref{j} we have already quantified the effect of
uncertainties in model parameters, QSO redshifts and cosmology. We
now discuss a number of other uncertainties associated with the
present study.

It is clear that our method is sensitive to errors in the continuum
placement, probably more so than the generic `line counting' method.
In \citeN{Liske00} we attempted to account for random errors in the
lowest orders of the continuum fit by determining the normalization of
the mean optical depth, $B$, for each spectrum separately. Here we
used a single value of $B$ derived from the entire dataset. This has
the advantage of making our estimate of $J$ fairly insensitive to the
adopted value of $\gamma$ (the evolutionary power law index). The
disadvantage is that errors in the continuum placement should, in
principle, increase the scatter when comparing the significance of the
proximity effect in individual lines of sight (cf.\ Fig.~\ref{mlmz}).
However, this comparison is based on measurements near the \lya\
emission lines of the QSOs where the S/N is in general quite high and
thus the continuum more secure. A {\em systematic} over- or
underestimation of continua can only affect our results if such a bias
is different for different parts of the spectra. This difference could
be caused by the higher S/N and greater curvature of the continuum in
the wing of the \lya\ emission line.

As our method deals only with the transmitted flux it circumvents all
problems that arise from defining a sample of individual absorption
lines. These problems include line blending, curve of growth effects
(e.g.\ \citeNP{Scott00}) and Malmquist bias \cite{Cooke97}.

The environment of QSOs may well differ from the intergalactic
environment in other aspects than just the intensity of ionizing
radiation. Most importantly, there may be additional absorption in the
vicinity of QSOs above and beyond the absorption already accounted for
in our model. If QSOs are hosted by groups or clusters of galaxies
then the gravitational pull of the host will cause infall of the
surrounding material and may thus increase the absorption line density
near a QSO \cite{Loeb95}. If this effect is not taken into account,
the proximity effect will appear weaker than it should and a larger
value of $J$ will be inferred. \citeN{Loeb95} suggested that the
magnitude of this effect may be as large as a factor of $\sim 3$. If
QSO luminosity is correlated with host mass, then more luminous QSOs
are affected more strongly by clustering and the value of $J$ derived
from the brightest QSOs should be higher than that derived from the
faintest. On the other hand, if all QSOs were affected by clustering
in the same way, then one might expect the observed slope of the
significance-luminosity relation in Fig.~\ref{mlmz} to be larger than
the one expected on the basis of the measured (and overestimated)
$J$. This is actually the case, although not significantly so.
However, all of the QSOs in the present sample are radio-quiet. The
hypothesis that such QSOs reside in rich galaxy cluster environments
has been repeatedly rejected at low redshifts and there seems to be
little evolution in the environment of radio-quiet QSOs up to $z <
1.5$ (\citeNP{Croom99}; \citeNP{Smith00}). In any case, it is most
likely very difficult to disentangle clustering from the proximity
effect and we believe that this issue deserves further study. For now,
it remains an uncertainty.

In Section~\ref{ionmodel} we assumed that the column density of an
absorber is proportional to the inverse of the incident ionizing
flux. This is only true for absorbers composed purely of hydrogen.
However, \citeN{Scott00} found that the inclusion of metals into the
model has an insignificant effect on the derived value of $J$. 

We have also ignored the fact that as radiation travels from a QSO to
a given absorber it will be attenuated by all the intervening
absorbers. In particular, an intervening strong Lyman limit system
will essentially `black out' the QSO entirely. In principle,
disregarding this effect causes overestimation of $J$ but
\citeN{Cooke97} concluded that the effect is negligible for $\log N <
17$. Higher column densities produce a conspicuous continuum break at
the Lyman limit. Unfortunately, only three of our spectra cover any
part of the Lyman limit region. One of these, Q0042--2656, shows a
Lyman break but the system lies $\sim 5200$\kms\ away from the
QSO. In any case, since high column density absorbers are
comparatively rare we would expect that only a fraction of our QSOs
are afflicted by this problem. However, from Fig.~\ref{mlmz} it is
apparent that none of our QSOs shows a proximity effect which is
unusually small for its luminosity and redshift. We thus find it
unlikely that we have significantly overestimated $J$ due to this
effect.

Obscuration by dust in intervening damped \lya\ absorption systems may
cause a QSO's luminosity and consequently $J$ to be underestimated
\cite{Srianand96}. Since there are no known damped \lya\ absorption
systems or candidates in our sample we believe that our value of $J$
is not affected by dust obscuration.

However, there are a number of other factors that create uncertainty
in the estimated luminosities of the QSOs. Errors in $K$-corrections
and the QSO continuum slope could be avoided by direct
spectrophotometric observations, but there are at least two other more
fundamental and probably larger uncertainties:

1. QSO variability. The equilibration time-scale of \lya\ absorbers,
$t_{\rm eq}$, is of the order $10^4$ years. Thus the observed
ionization state of an absorber in the vicinity of a QSO will
approximately reflect the ionizing flux received from that QSO
averaged over the $10^4$ years prior to the epoch of observation,
$\bar L$. Let us assume that the intrinsic luminosity of QSOs varies
on a single time-scale, $t_{\rm V}$. For $t_{\rm V} \la t_{\rm eq}$
the QSO's observed luminosity may be different from $\bar L$ and thus
the strength of the observed proximity effect may be different from
that expected on the basis of the ionization model. The same may be
true for the foreground proximity effect even when $t_{\rm V} > t_{\rm
eq}$ if $t_{\rm V}$ is smaller than the light travel time from the IS
to the observed absorption ($\approx 10^6$~years). Obviously, we have
no information on QSO variability on such large time-scales. However,
\citeN{Giveon99} found that on time-scales of $100$ to $1000$ days the
distribution of brightness deviations about the mean light curves of
$42$ QSOs has a width of $0.14$~mag in the $B$-band. Thus, variability
on short time-scales contributes substantially to the uncertainty in
the QSOs' luminosities.

2. Gravitational lensing. None of the QSOs in our sample are known to
be lensed and at least four of them have been included in searches for
multiply imaged QSOs with negative results \cite{Surdej93}, implying
that they are at least not strongly lensed. However, some or all of
them may be weakly lensed by the non-homogeneous distribution of
foreground matter on large scales. There are two known overdensities
of \ion{C}{iv} absorption systems in the foreground of the present QSO
sample \cite{Williger96} which could possibly cause weak lensing
(\citeNP{VandenBerk96}; \citeNP{Holz98}). Ray-tracing experiments
performed on simulations of cosmological structure formation have
yielded the probability distribution function of the magnification
caused by galaxy clusters and large-scale structure
\cite{Hamana00,Wambsganss98}. At $z = 3$ the dispersion of this
distribution can be as high as $0.4$ (with a mean of $1$) and the
distribution has a power law tail towards large magnification. For the
standard model the probability of encountering a magnification of $2$
or greater is a few per cent but for a $\Lambda$ model it is less than
$10^{-3}$. A magnification by a factor of $1.2$ (or a demagnification
by a factor $1.2^{-1}$) introduces an additional uncertainty of $\pm
2.5 \log 1.2 = \pm 0.2$~mag which is actually larger than the quoted
measurement errors on the observed $B$-band magnitudes.

What are the effects of luminosity uncertainties on our results if we
treat them as {\em random} errors? Any statistical errors should
increase the error on $J$. We can estimate the magnitude of this
effect by the following argument. If we measured $J$ from a single QSO
then the additional error on $J$ should be on the order of $\Delta
(\log J) \approx 0.4 \Delta m$, where $\Delta m$ is the error on the
QSO's magnitude. For ten QSOs this error should be smaller by a factor
of $\sqrt{10}$. Thus for $\Delta m = 0.5$~mag we find $\Delta(\log J)
= 0.063$ which is much smaller than the quoted error of $0.4$.

This also shows that the {\em known} QSO variability, gravitational
lensing and measurement errors on the QSO magnitudes cannot increase
the errorbars on our foreground and background estimates of $J$ by an
amount large enough to make these two values compatible.

However, statistical errors in $L_\nu$ will weaken our results on the
correlation between the significance of the proximity effect and
$L_\nu$. Clearly, for errors as large as $\Delta m = 0.5$~mag (which
corresponds to a factor of $1.6$ in luminosity) fitting a straight
line to the data points in Fig.~\ref{mlmz} will be almost
meaningless. In Section~\ref{lumzcor} we estimated that the
significance of the proximity effect is correlated with luminosity at
the $2.6\sigma$ level. For $\Delta m = 0.3$~mag this significance
drops to $2.1\sigma$. Thus measurement errors and the known
variability of QSOs alone cannot entirely invalidate the evidence for
a correlation of the background proximity effect with QSO Lyman limit
luminosity.

In any case it is probably more appropriate to treat both QSO
variability and gravitational lensing as {\em systematic} errors.
Because of the steepness of the bright end of the QSO luminosity
function a magnitude limited sample of QSOs is more likely to contain
magnified QSOs than demagnified ones (e.g.\ \citeNP{Pei95};
\citeNP{Hamana00}). The same is true for QSOs that are near a peak in
their lightcurves (e.g.\ \citeNP{Francis96}). Thus either of these
effects could cause a systematic overestimation of $J$. Together they
imply that on average the QSO magnitudes may have been overestimated
by $\sim 0.35$~mag which corresponds to an overestimation of $J$ by
$\Delta(\log J) = 0.14$.

Note that gravitational lensing and QSO variability on short
time-scales ($ < 10^4$ years) affect our $J$ estimates from the
classical and foreground proximity effects in the same way. However,
if QSO luminosities vary on time-scales of $\sim 10^6$ years then this
will affect only the latter estimate as explained above. In this case
we would expect the QSOs to have been systematically fainter over a
period of $\sim 10^6$ years prior to the time they emitted the photons
which we receive today, causing us to overestimate $J$ when measuring
it from the foreground proximity effect. To reconcile the foreground
and background values of $J$ we require a variability of $\Delta m =
2.5 \log 2.2 = 0.86$~mag (assuming $\Delta v = 800$\kms) on
time-scales of $\sim 10^6$ years.

Past authors \cite{Cooke97,Scott00} have attempted to test for the
presence of gravitational lensing in their data: since high luminosity
QSOs are more likely to be lensed than low luminosity ones, an
estimate of $J$ from the former group should be higher than from the
latter. From Fig.~\ref{mlmz} we can see that the higher luminosity
QSOs of our sample actually show a slightly more prominent proximity
effect than expected for $\log J_{21} = -0.45$. Thus a $J$ measurement
from these four QSOs will yield a lower value, contrary to the
expectation if they were lensed.

\section{Conclusions} \label{conclusions}
We have analysed the \lya\ forest spectra of a close group of $10$
QSOs in search of the (foreground) proximity effect using a novel
method based on the statistics of the transmitted flux. We list our
various measurements of $J$ in Table~\ref{jtab} and we summarise our
main results as follows:

1. We confirm the existence of the classical background proximity
   effect at the $> 99$ per cent confidence level.

2. From the observed underdensity of absorption near the background
   QSOs we derive $\log J_{21} = 0.1^{+0.4}_{-0.3}$ (90 per cent
   confidence limits).

3. Correcting all QSO redshifts by $\Delta v = +800$\kms\ we find
   $\log J_{21} = -0.45^{+0.4}_{-0.3}$. The reduction of $J$ with
   $\Delta v$ is consistent with previous results.

4. The significance of the background proximity effect in individual
   lines of sight is correlated with QSO Lyman limit luminosity at the
   $2.6\sigma$ level, thus lending further support to the hypothesis
   that the proximity effect is caused by the additional UV flux from
   background QSOs. We account for the fact that the significance is
   also correlated with redshift which is due to the evolution of the
   absorption line density. We considered an alternative model for the
   proximity effect where the underdense absorption is caused by some
   hypothetical property $x$ of the QSO or its environment. We find
   that the property $x$ would have to be noticeably correlated with
   luminosity in our sample and probably in the QSO population in
   general.

5. The full sample shows no evidence for the existence of the
   foreground proximity effect.

6. This absence implies a lower limit of $\log J_{21} > 0.3$. If we
   interpret the discrepancy of this lower limit with previous
   measurements as evidence that QSOs radiate anisotropically, then
   they must emit at least a factor of $1.4$ less ionizing radiation
   in the plane of the sky than along the line of sight to Earth.

7. Our sample includes the fortunate constellation of a foreground QSO
   surrounded on all sides by four background QSOs with approximately
   equal separations from the foreground QSO. Contrary to the rest of
   the sample, this particular QSO induces a foreground proximity
   effect in the surrounding lines of sight at the $3.5\sigma$
   level. For this subsample we measure $\log J_{21} =
   -0.65^{+0.6}_{-0.5}$.

8. Finally, we have discussed possible sources of systematic and
   additional statistical errors. We conclude that clustering of
   absorption systems around QSOs is the most likely source of
   systematic error. We also find that the known variability of QSOs
   reduces the significance of the proximity effect-luminosity
   correlation to $2.1\sigma$.
	
\begin{table}
\caption{Summary of $J$ measurements and limits.}
\label{jtab}
\begin{tabular}{lrc}
\hline 
& $\log J_{21}$ & Errors$^a$\\ 
\hline
standard & $0.10$ & $+0.40, -0.30$ \\
$\beta = 1.3$ & $-0.45$ & $+0.45, -0.40$ \\
$\beta = 1.7$ & $0.60$ & $+0.30, -0.25$ \\
$\gamma = 2.5$ & $0.15$ & $+0.35, -0.30$ \\
$q_0 = 0.15$ & $0.15$ & $+0.40, -0.30$ \\
$\Delta v = 400$\kms & $-0.25$ & $+0.40, -0.30$ \\
$\Delta v = 800$\kms & $-0.45$ & $+0.40, -0.30$ \\
foreground & $> 0.30$& \\
Q0042--2639 quadrangle & $-0.65$ & $+0.60, -0.50$ \\  
\hline
\end{tabular}\\
$^a$Errors are $90$ per cent confidence limits.\\
\end{table}

From Fig.~\ref{prev_j} we can see that our measurement of $J$ from the
background proximity effect is consistent with most previous
measurements. \citeN{Bechtold94} found a somewhat larger value $\log
J_{21} = 0.5$ but she did not correct the QSO redshifts, remarking
only that a correction of $\Delta v = 1000$\kms\ would lower her
value of $J$ by a factor of $3$. \citeN{Srianand96} also found a
considerably higher value which is partly due to the fact that they
used $\beta = 1.7$. \citeANP{Kulkarni93}'s \citeyear{Kulkarni93} low
value of $J_{21} = -2.2$ at $z \approx 0.5$ is usually taken as
evidence for an evolving background.

\citeN{Bechtold94} and \citeN{Scott00} noted that the relative deficit
of absorption lines near background QSOs was larger for the
high-luminosity halves of their samples than for the low-luminosity
ones. However, they did not quantify this effect in any detail and did
not compare it to the expectations from the ionization model. BDO
identified a trend of the line deficit with luminosity at $1\sigma$
significance. \citeN{Lu91} on the other hand found no such trend. In
Figs.~\ref{mlmz}, \ref{mlz} and \ref{pmlz} we have presented good
evidence that the significance of the proximity effect is indeed
correlated with QSO luminosity and redshift. This result was possible
because we used a technique that is more sensitive to variations of
the absorption density on large scales than the generic line counting
method. Using simulations we find that these correlations are entirely
consistent with the expectations of the ionization model. However,
considering the discussion of Section~\ref{uncertainties} we clearly
need to apply our method to a larger sample with a wider range in
luminosity to establish this result more firmly. Nevertheless, the
present analysis provides further evidence that the interpretation of
the proximity effect as being due to increased ionization caused by
the extra UV flux from the background QSO is essentially correct.

Although their results were poorly constrained, it is interesting to
note that the three BQSO-IS pairs of \citeN{Fernandez95} favoured a
similarly high value of $J$ as our full sample of $14$ pairs. The
non-detections of the foreground proximity effect by \citeN{Crotts89}
and \citeN{Moller92} also imply a high value of $J$. On the other
hand, all possible positive detections of the foreground proximity
effect (\citeNP{Dobrzycki91}; \citeNP{Srianand97}; the Q0042--2639
quadrangle) yield $J$ values that are in line with the measurements
from the background effect.

One possible explanation is that most of the QSOs were substantially
fainter over a period of $\sim 10^6$ years prior to the time they
emitted the photons which we receive today. This would cause an
overestimate of $J$ when measured from the foreground proximity effect
but would not affect the results of the background proximity effect.

As we have already pointed out, the other explanation is that QSOs
radiate anisotropically. There is a large body of observational
evidence which suggests that the observed characteristics of an Active
Galactic Nucleus (AGN) depend on the direction from which it is viewed
(see e.g.\ \citeNP{Antonucci93} for a review). The basic theme of
unified models for AGN is that some or even all of the many different
types of AGN are in fact the same type of object but seen from
different directions. In these models the directionality is caused by
a thick, dusty and opaque torus which surrounds the central
engine. The smooth continuum and broad emission lines of a QSO are
thought to originate from within the torus and thus they can only be
seen indirectly by scattered light when the torus is viewed
approximately edge on.

This scenario has two obvious implications: i) all QSOs show a
background proximity effect and ii) whether a foreground proximity
effect is seen or not depends on whether nearby absorption systems
probed by other sightlines can `see' inside the torus of the QSO. If
so, they will roughly see the same continuum as we do, resulting in a
measurable depletion of absorption and a corresponding $J$ value which
is similar to that derived from the background effect. If not, there
will be little or no foreground effect, resulting in a high value or
lower limit for $J$.

Assuming a simple picture of this kind, \citeN{Dobrzycki91} used the
velocity offset of their void from the foreground QSO to derive a
value of $\sim 140$\degr\ for the opening angle of the torus.

However, it is difficult to explain the case of the Q0042--2639
quadrangle with this scenario. If all of the four underdensities are
real and caused by the forground QSO then it has to emit similar
amounts of radiation along three nearly perpendicular axes
(north-south, east-west and towards Earth), leaving little room for
anisotropic emission.

The current situation is thus uncertain and intriguing enough to
stimulate further observations. The motivations and potential gain are
clear: by investigating the radiative effects of QSOs (or other AGN)
on nearby absorption systems along other lines of sight we can `view'
them from different directions which may help to constrain unified
models of AGN. The ideal targets for further observations are dense
groups of QSOs at similar redshifts. These are rare in current QSO
catalogues but, fortunately, currently ongoing redshift surveys
like 2QZ will soon remedy this situation.

\section*{Acknowledgments}
We thank J.~Baldwin, C.~Hazard, R.~McMahon and A.~Smette for kindly
providing access to the data. JL acknowledges support from the German
Academic Exchange Service (DAAD) in the form of a PhD scholarship.

\label{lastpage}

\end{document}